\documentclass[aps,pre,showpacs,twocolumn]{revtex4}
\usepackage{amssymb}
\usepackage{graphicx}
\usepackage{colordvi}
\usepackage{color}
\usepackage{dcolumn,amsmath,amsthm,amscd,amsfonts,amssymb,epsfig,graphics,graphicx,eucal}

\newcommand{\be}{\begin{equation}}
\newcommand{\ee}{\end{equation}}

\begin{document}

\title{Coupling between Fano and Bragg  bands in photonic band structure\\ of two-dimensional metallic photonic structures }
\author{P. Marko\v s$^{1}$ and  V. Kuzmiak$^2$}
\affiliation{ $^1$Department of Experimental Physics, Faculty of  Mathematics Physics and Informatics, Comenius University in Bratislava, 842 28 Slovakia
\\
$^2$  Institute of Photonics and Electronics, Academy of Sciences of the Czech Republic,v.v.i., Chaberska 57, 182 51, Praha 8, Czech Republic
}

\date{\today}

\begin{abstract}
Frequency and transmission spectrum of two-dimensional array of metallic rods is investigated numerically. Based on the recent analysis of the   band structure of two-dimensional photonic crystal with dielectric rods [P. Marko\v{s}, Phys. Rev. A 92 043814  (2015)]  we identify two types of bands in the frequency spectrum: Bragg ($\mathcal{P}$) bands resulting from a periodicity and Fano ($\mathcal{F}$) bands which arise from Fano resonances associated with each of the cylinders within the periodic structure. It is shown that the existence of Fano band in a certain frequency range is manifested by a Fano resonance in the transmittance. In particular, we re-examine the symmetry properties of the $H$-polarized band structure in the frequency range where the spectrum consists of the localized modes associated with the single scatterer resonances and we explore process of formation of Fano bands by identifying individual terms in the expansion of the  LCAO states. We demonstrate how the interplay between the two scattering mechanisms affects properties of the resulting band structure when the radius of cylinders is increased. We show that a different character of both kinds of  bands is reflected in the spatial distribution of the magnetic field which displays patterns corresponding to the corresponding irreducible symmetry representations.
\end{abstract}
\maketitle

\input{epsf.tex}
\epsfverbosetrue

\section{Introduction}
\label{intro}

The properties of photons in photonic crystal can be described efficiently in terms of band structure. When the dielectric constants of materials which compose the photonic crystals are sufficiently different and the absorption of light by the materials is sufficiently low, the photonic band structure might exhibit a photonic gap in which photons are forbidden to propagate in certain directions with specified frequencies \cite{joan-pc,sakoda,costas}. The physical origin of separated frequency bands in photonic crystals can be described in terms of two scattering mechanisms. The first one, the Bragg-like scattering \cite{sakoda} may lead to existence of a gap in the total photon density of states due to the destructive interference of the scattered electromagnetic (EM)  waves from a periodic array of resonators and can give rise large scale or macroscopic resonance. The second mechanism which strongly influences the frequency spectrum of photonic crystals, is based on microscopic resonances associated with independent, uncorrelated scatterers. This scattering plays a key role in the theory of disordered photonic structures, where the interference of scattered waves may cause the spatial localization \cite{sjohn1} of the EM wave when the density of resonant centers possess the value for which spheres of the influence of the scatterers become optically connected \cite{sjohn2}. The latter mechanism can be described in terms of the free-photon Ioffe-Regel criterion of localization \cite{ioffe,meso} $2\pi \ell/\lambda \simeq 1$, where $\ell$ is the mean free path and $\lambda$ is the vacuum wavelength. According to this criterion a transition from extended to localized modes takes place due to the interference of multiple scattering paths \cite{sjohn1,sjohn2,meso}. When the size of scatterers becomes comparable to the wavelength $\lambda$, Mie resonances \cite{mie,Hulst} are excited and consequently affect the propagation of the EM waves. For example, it was demonstrated that that there is a direct correspondence between the gaps calculated by plane-wave expansion method \cite{economou,kafesaki} and the Mie resonances. A large photonic band gap arises due to the synergetic interplay between the microscopic and macroscopic resonance mechanisms, i.e. when the density of dielectric spheres is chosen such that the Mie-resonances in the scattering by individual scatterers occur at the same wavelength as the macroscopic Bragg resonance of an array of the same scatterers \cite{sjohn1}.

On the other hand, when the concentration of the scatterers increases, the excitation of the resonances enables the propagation of the EM wave across the sample. In this case Mie resonances excited at isolated scatterers play the role of the electron eigenstates  associated with isolated atoms in a crystal or spatially localized states in disordered structures \cite{pendry}. Note that in contrast to Bragg scattering, Mie scattering does not require a perfect periodic arrangement of scatterers.

In dealing with this analogy, however,  one has to take into account two following important differences.
At first, in contrast to the case of electronic structure, the photons unlike electrons are not bound to a single scatterer and transmission is primarily achieved through  the scattering resonances which occur when the wavelength of light is comparable to the size of the scatterer.
At second, in the case of classical wave  the medium supports propagating solutions for every frequency and thus for large wavelengths the transmission occurs due to this propagation mode even when no localized resonances are excited. These assumptions have been  theoretically implemented in tight-binding formulation of light propagation in two-dimensional photonic band gap structures \cite{Lidorikis}, and in particular, the role of single scatterer resonances in formation of the higher frequency bands was verified.

The recent interest in the investigation of the modes with a complex wave vector \cite{davanco,notaros,rybin2016} has led to significant advancement in the understanding of both scattering mechanisms which manifests themselves in the resulting complex band structures. In particular, both alternative (inverse dispersion approach) \cite{rybin2016}  and revisited forms of the conventional computational techniques(FEM,finite-difference method) \cite{davanco,notaros} inherently yield bands of pure imaginary and complex-wave vector Bloch modes. It has been suggested  that the modes with pure imaginary wave vector which exist within the band gaps play an important role in representing the evanescent field inside
a finite or semi-infinite photonic crystal slab under external excitation \cite{istrate}. The properties of the modes with complex wave vector proved to be even more intrigue as they can exist both within and outside the band gaps. Specifically, they significantly affect the transmittance and reflectance in the subwavelength plasmonic crystals \cite{davanco} and give rise to Fano-like resonances. On the other hand, the complicated features originating from Bragg or Mie evanescent modes that are found in the complex band structures of an infinite two-dimensional photonic crystals allow to discern unambiguously between Bragg and Mie gaps in the spectra \cite{rybin2016}.

Curiously enough until recently \cite{rr,rybin,rybin2015}, both mechanisms -- coherent Bragg scattering and transmission via microscopic resonances on scatterers --  were to our knowledge considered to take part independently in the formation of the lower and higher frequency bands, while their coupling has not been systematically studied. The relation between these two mechanisms which unveils new effects associated with their synergetic interplay has been investigated on the example of a finite two-dimensional photonic crystal consisting of infinitely long dielectric cylinders \cite{PM1}. It was shown that the bands which form photonic band structure of 2D photonic crystals consists of two types of bands  --
 $\mathcal{P}$ and  $\mathcal{F}$ bands --
which arise either (a) due to the permittivity contrast between the cylinders and a background which lead to the formation of gaps at the edges of the first Brillouin zone
 ($\mathcal{P}$  bands) \cite{fP}
 or  (b) from Fano resonances associated with each of the cylinder which form the bands when are brought together to form 2D periodic array
 ($\mathcal{F}$  bands).  The different nature of both types of bands is reflected in the spatial distribution of the electric field in dielectric cylinder. The bands are classified in terms of the irreducible representations of $C_{4v}$ symmetry group. Their properties are determined by the symmetry of the terms in the expansions of LCAO states and by the order of excited Fano resonance \cite{Fano,sfan}. The coupling between the periodic $\mathcal{P}$ bands and Fano $\mathcal{F}$ bands give rise to irregularities in the resulting spectrum such is splitting of the $\mathcal{P}$ band or overlapping between them.

The casting of the bands described above provides a convenient approach that can be applied to describe $H$-polarized photonic band structure of 2D photonic crystals consisting of metallic rods. The origin of nearly dispersionless flat bands in the $H$-polarized spectrum of metallic rods has been identified as the weak overlap of $H$-polarized excitations associated with each metallic cylinder\cite{Economou} that are characterized by discrete frequencies\cite{KMP}. When a sufficiently large or infinite number of cylinders is brought together to form a periodic structure the overlap of these excitations broadens the discrete frequencies into narrow bands. The physical origin of these nearly dispersionless bands has been confirmed by subsequent investigations employing the finite-difference time-domain (FDTD) technique\cite{KS} and the multiple multipole method(MMP) \cite{EM} capable to deal with surface modes which proved to be more appropriate to calculate band structures with components characterized by frequency-dependent dielectric function than the plane-wave methods \cite{KMP}. There were identified two kinds of modes within the resulting $H$-polarized spectrum: (a) modified plane waves which likewise $E$-polarized bands have their replicas in free space and (b) resonant states which can be regarded as analogue of linear combination of atomic orbitals in solid state theory and which  are associated with localized plasmon resonances localized close to the surface of cylinders and their symmetry can be classified  in terms of the resonant states corresponding to the Mie resonances of an isolated cylinder \cite{KS,EM,Vala}.

In our paper we apply scattering approach which is based on the expansion of electromagnetic field in the series  of the cylinder functions \cite{stratton}.
In contrast to results reported previously \cite{KS,EM} we calculate transmission of EM wave of the 2D array of cylinders which is finite in the direction of propagation.
This technique allows us to study  spectral properties of structures starting from an infinite  one-dimensional array of cylinders which is consequently extended to the form the two-dimensional photonic crystal. This approach proved to be feasible in the case of two-dimensional  array of  dielectric rods and it allows investigating a coupling between localized and extended bands as well reveals new features associated with a finite size of the structure.

\section{Structure and numerical method}
\label{method}

\begin{figure}[t]
\noindent\includegraphics[width=0.3\textwidth]{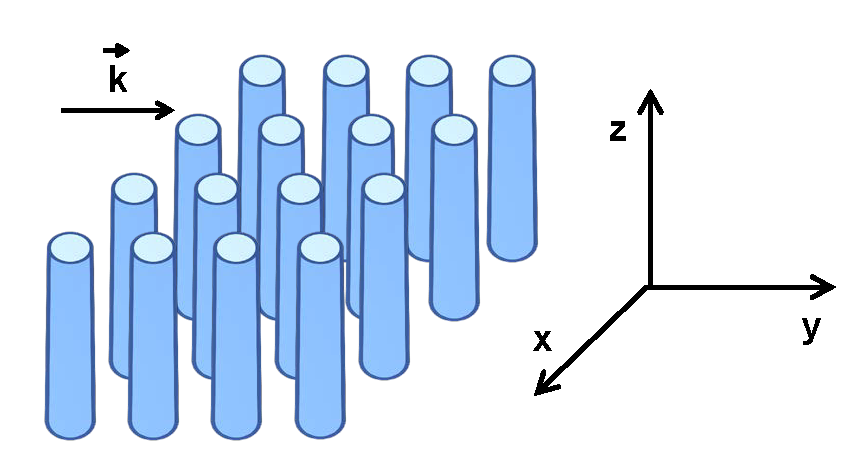}
\caption{Two-dimensional periodic array of metallic rods.
The structure is infinite in the $x$ direction. The number $N$ of cylinders in the $y$ direction varies from $N=1$ (linear array of rods) to $N=24$.
}
\label{fig-sample}
\end{figure}

The system we consider in this paper consists of a finite number of linear arrays of metallic cylinders of radius $R$
and  of infinite length along the $z$ axis which form a finite   two-dimensional photonic slab constructed typically from $N$
rows of the same cylinders located in planes $y = n_ya$, $n_y = 0,1,\dots, N-1$ (Fig.~\ref{fig-sample}).
The dielectric function of metal from which
cylinders are  formed is assumed to have the simple, free-electron form
\begin{equation}
\epsilon(\omega) = 1 - \omega_p^2/\omega^2,
\label{Drude}
\end{equation}
 where $\omega_p$ is the plasma frequency of the conduction electrons. The frequencies in the results for the band structures
 and the transmittances are normalized to the plasma frequency $\omega_p$ by setting $\lambda_p = a$, where $\lambda_p$ is the
 plasma wavelength.

The electromagnetic wave is assumed to propagate in the $xy$ plane, perpendicular to the rods. The  incident EM wave
\be
H_z(x, y |\omega)_{\rm inc} =  \exp[i (k_x x+  k_y y) - i\omega t]
\ee
is polarized parallel to the axes of the cylinders. To describe the properties of EM waves in a two-dimensional photonic crystal
we calculate the transmission coefficient  as a function of frequency and evaluate the spatial distribution of magnetic field.
The algorithm which we use is based on the expansion of electromagnetic field into cylinder functions \cite{Hulst,stratton,Ohtaka1,Ohtaka2,asa,natarov}.
Our approach is described in detail in Ref. \onlinecite{PM3}. The magnetic field scattered at a cylinder centered at $\vec{r} =(0,0)$  can be
expressed in cylindrical coordinates $r$ and $\phi$ as
\begin{equation}
H^{\rm in}_z  = \alpha^+_0\tilde{J}_0 + 2\sum_{k>0}\alpha^+_k\tilde{J}_k \cos(k\phi)
 + 2i\sum_{k>0}\alpha^-_k\tilde{J}_k \sin(k\phi)
\label{eq.1a}
\end{equation}
and
\begin{equation}
H^{\rm out}_z  = \beta^+_0\tilde{H}_0 + 2\sum_{k>0}\beta^+_k\tilde{H}_k \cos(k\phi)
 + 2i\sum_{k>0}\beta^-_k\tilde{H}_k \sin(k\phi)
\label{eq.1b}
\end{equation}
for $r < R$ and $r > R$, respectively.  Here, $\tilde{J}_k (r) = J_k (2\pi rn/\lambda)$,
$\tilde{H_k} (r) = H_k (2\pi r/\lambda)/H'_k (2\pi R/\lambda)$, $J_k (r)$ are the Bessel
functions and $H_k (r)$ and $H'_k (r)$ are the Hankel functions of the first kind and their derivatives \cite{note1}, respectively.
 $\lambda = 2\pi c/\omega$ is the wavelength of the EM field in the vacuum and $n = \sqrt{\epsilon \mu}$ is the refraction index. To evaluate the
 field $H_z$ scattered by a cylinder centered at $\vec{r}_{n_x n_y} = (n_x a,n_y a)$  one can use the Eqs.(\ref{eq.1a}) and (\ref{eq.1b})
 with a new set of coefficients $\alpha_k(n_x,n_y)$ and  $\beta_k(n_x,n_y)$ and cylindrical coordinates $r$ and $\phi$ corresponding to the position of the cylinder.

 The coefficients $\alpha_k(n_x,n_y)$ and  $\beta_k(n_x,n_y)$ can be calculated from the  continuity condition  of the tangential components of the electric and magnetic
 field at the boundary of cylinders. Note that two components of the electric field $E_r$ and $E_{\phi}$ can be expressed by using of the same set of
 coefficients $\alpha$ and $\beta$ by using the expressions
\be
E_r = \frac{i}{\omega\epsilon r}\frac{\partial H_z}{\partial \phi},
~~~~~
E_\phi =  -\frac{i}{\omega\epsilon }\frac{\partial H_z}{\partial r}.
\ee

The spatial periodicity of the structure along the $x$-axis allows us to reduce significantly the  number of
unknown coefficients since the coefficients $\alpha(n_x,n_y)$ and  $\beta_(n_x,n_y)$ satisfy the Bloch theorem
\begin{eqnarray}
\alpha_k(n_x,n_y) = \alpha_k(0,n_y)e^{ik_x an_x} \\
\beta_k(n_x,n_y) = \beta_k(0,n_y)e^{ik_x an_x},
\label{Bloch}
\end{eqnarray}
By employing the relation given by Eq.\ref{Bloch} the number of unknown coefficients $\beta^{\pm}_k (n_y)$  is reduced to $N\times(2N_B+1)$, where $N_B$ is the highest order of  the Bessel function. The transmission coefficient
\begin{eqnarray}
T = \frac{S_y(y_p)}{ S^i_y} = \frac{\int^{a/2}_{-a/2}dx e_x(x,y_p)h^*_z(x,y_p)}{\int^{a/2}_{-a/2}dx e^{i}_{x}(x,-y_p)(h^{i}_{z}(x,-y_p))^*}
\label{Poynting}
\end{eqnarray}
at the opposite side of the structure can be calculated  as the ratio of the $y$-component of the Poynting vector $S_y(y_p)$, to the incident Poynting vector
$S^i_y$ for any $y_p>Na$ \cite{PM3}.
The band spectrum can be calculated from the following $y$-dependence of coefficients $\beta$
\be
\beta_k(0,n_y) = \beta_k(0,0) e^{iqan_y}
\label{eq.q}
\ee
which determines the wave vector $q$ \cite{note3}.

As we show in the next section, incident EM wave excites in cylinders resonances which could be 
identified from resonant behavior of coefficients  $\beta_k$. The order of the resonance is given by index $k$. 
For an isolated cylinder the
resonances are shown in  Fig. \ref{H-cylinders}. In a linear array of cylinders, each resonance splits 
into two ones -- even and odd with respect to the  symmetry of EM field along the $x$ direction (Figure \ref{koeficienty}). 
(Note that the frequency of even resonance lies below the odd resonant frequency.)  We argue that each 
observed resonance gives rise to the  Fano band in the 2D array of cylinders and identify these bands by number of Fano resonance and by its symmetry: $n\pm$. In our notation, bands $1-$, $2+$, $3-$ \dots are even  while $1+$, $2-$, $3+$, etc. are odd bands.

We calculate numerically the coefficients $\alpha$ and $\beta$ for 
incident plane wave and various configurations of metallic cylinders. 
For  2D array of cylinders shown in Fig. \ref{fig-sample} we obtain the transmission coefficients. 
(Eq.  \ref{Poynting}) and  
identify transmission bands shown in left panels in Figs. 3-5. Equation \ref{eq.q} enables us to identify the 
dispersion relation $\omega=\omega(q)$ and to construct the frequency band structure. 

The  symmetry of excited  modes in the photonic structure is even or odd with respect to the direction of  propagation of the incident wave at zero-incident angle. Therefore, to see  modes with odd symmetry
visible in the transmittance 
  a non-zero incident angle --
typically $\theta = \pi/100$  -- 
has to be applied in order to achieve a sufficient coupling with the  incident wave.

\section{results}
\label{results}

\subsection{An isolated cylinder}

We first study the resonances associated with an isolated cylinder which are excited by an incident plane wave \cite{Economou}.  We list resonant frequencies of
a single metallic cylinder of radius $R=0.3a$ in the first column of  Table \ref{table1}. The higher order resonances are associated with excitation of the surface plasmon.
Indeed, as the wavelength of the surface plasmon decreases when $\omega \rightarrow \omega_s = \omega_p/\sqrt{2}$  and becomes much smaller than the
radius $R$,  $\lambda \ll R$, then the surface of the cylinders acts nearly as a planar interface.  By assuming the oscillation along the circumference of the cylinder
one obtains from the cyclical boundary condition for the resonant wavelength $\lambda_n$  the condition
\begin{equation}
2\pi R = n\lambda_n.
\end{equation}
By substituting $\lambda_n$ into the expression for the wavelength of the surface plasmon one obtains
\begin{eqnarray}
\lambda_n = \lambda_p\sqrt{\frac{2\omega_n^2 - 1}{\omega_n^2 -1}}
\label{lambda_n}
\end{eqnarray}
which yields the resonant frequencies $\omega_n$ which agree very well with numerical results.

Two other columns of Table \ref{table1} present resonant frequencies found at 1D lattice of metallic cylinders.
Due to its symmetry, each resonance associated with 1D array splits into two separate resonances - even and odd.
It shown below that this splitting is responsible for the appearance of doublet frequency bands in the spectrum of
the 2D photonic crystal.

\begin{table}[t]
\begin{tabular}{|r|r|r|r|}
\hline
n   & ~~Cylinder~~ & ~~Chain: Even~~   & ~~Chain:  Odd~~  \\
\hline
\hline
2 &  0.61512  &  0.579  &0.65136 \\
3 &  0.66287  &  0.6568 & 0.6592 \\
4 &  0.68470 &    0.6892& 0.6826  \\
5 &  0.69365   &  0.69296&  0.69398  \\
6  &  0.69799 &  0.69831 & 0.69778 \\
7  & 0.7005 &  0.70043 & 0.70056 \\
8  & 0.70209 &  0.70211  & 0.70207 \\
9  & 0.70316  &  0.70315  & 0.70317\\
10 & 0.70392  &  0.70392  & 0.70392\\
11 & 0.70448  &  0.70448 & 0.70448\\
12 & 0.7049   &  0.7049  &  0.7049 \\
\hline
\end{tabular}
\caption{Resonant frequencies of an isolated metallic cylinder of radius $R=0.3a$  (Fig.~\ref{H-cylinders}) and for linear chain
of cylinders  (Fig.~\ref{koeficienty}). Even resonances correspond to coefficients $\alpha_{2k}^+$ and $\alpha_{2k-1}^-$,
according to the notation used in Eqs. (\ref{eq.1a}) and (\ref{eq.1b}).
}
\label{table1}
\end{table}

\begin{figure}[t]
\noindent\includegraphics[width=0.32\textwidth]{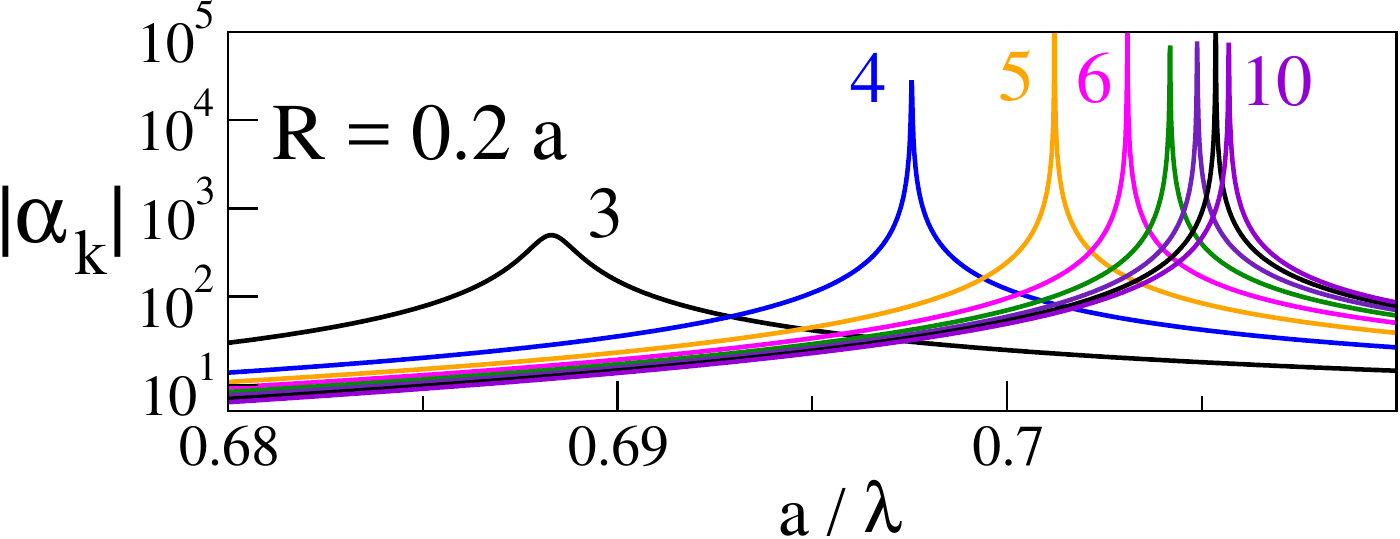}\\
   \includegraphics[width=0.32\textwidth]{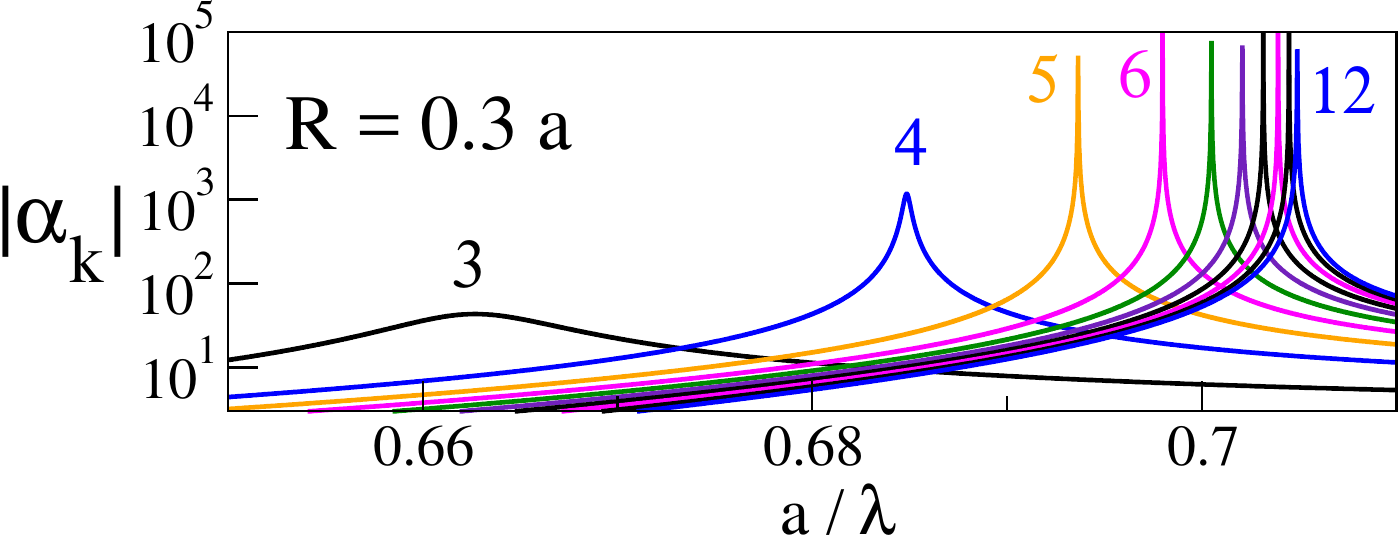}\\
          \includegraphics[width=0.32\textwidth]{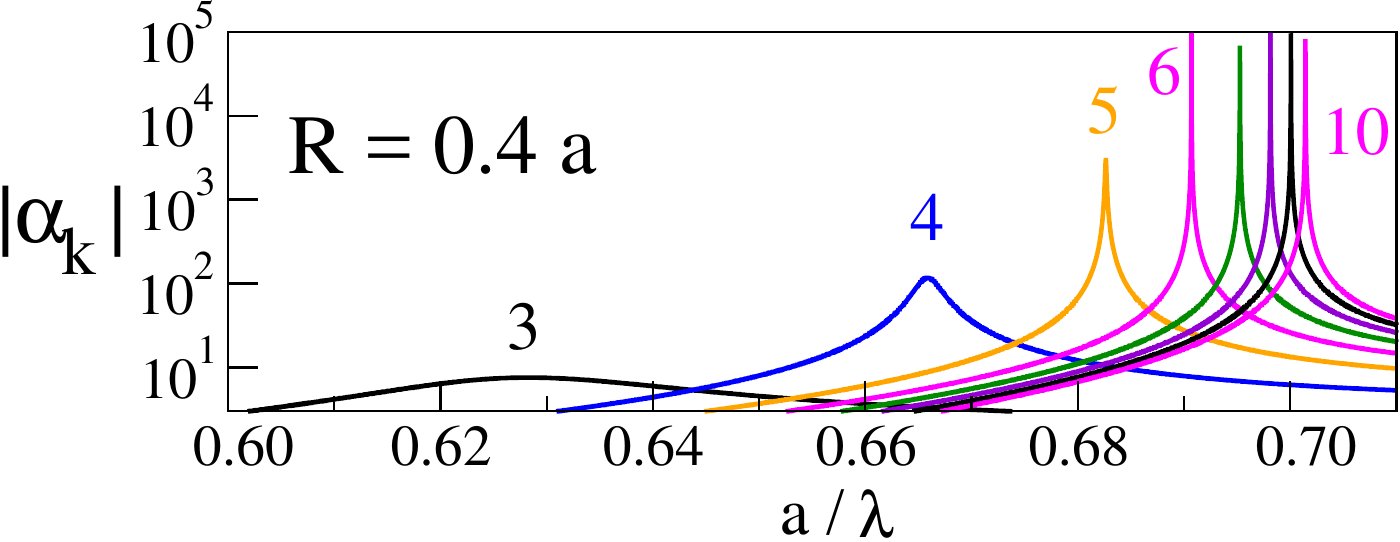}\\
\caption{(Color online) Mie resonances excited at metallic cylinders of various radius.
The order of resonances increases from the left to the right.}
\label{H-cylinders}
\end{figure}

\begin{figure}[t]
\noindent\includegraphics[width=0.23\textwidth]{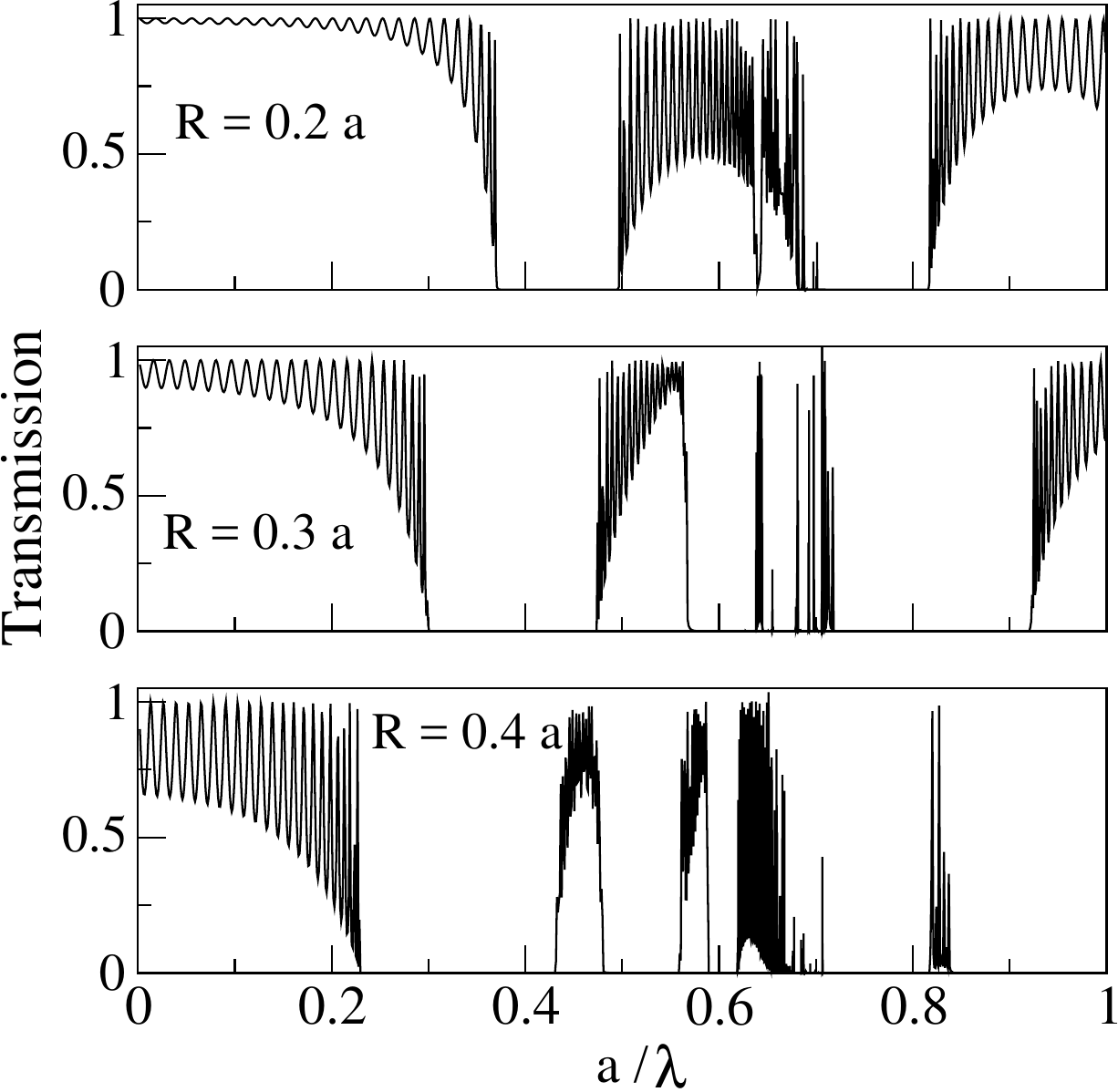}
~~
\noindent\includegraphics[width=0.23\textwidth]{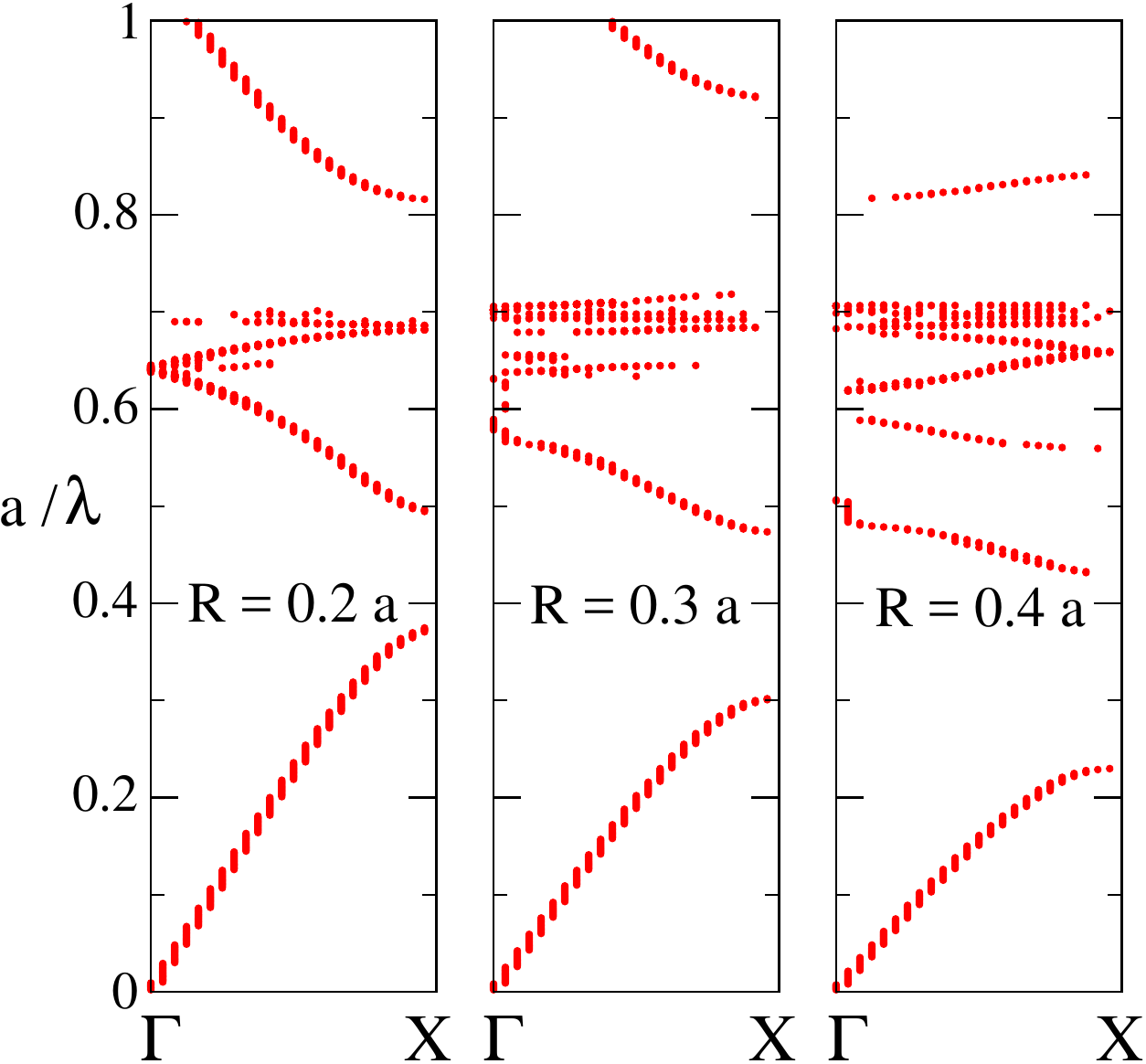}
\caption{Transmission coefficient (left) and band structure calculated from transmission data
for $N=24$ rows of cylinders in the frequency range $0<f<1.$
}
\label{Hr_02_03_04}
\end{figure}

\begin{figure}[h!]
\includegraphics[width=0.23\textwidth]{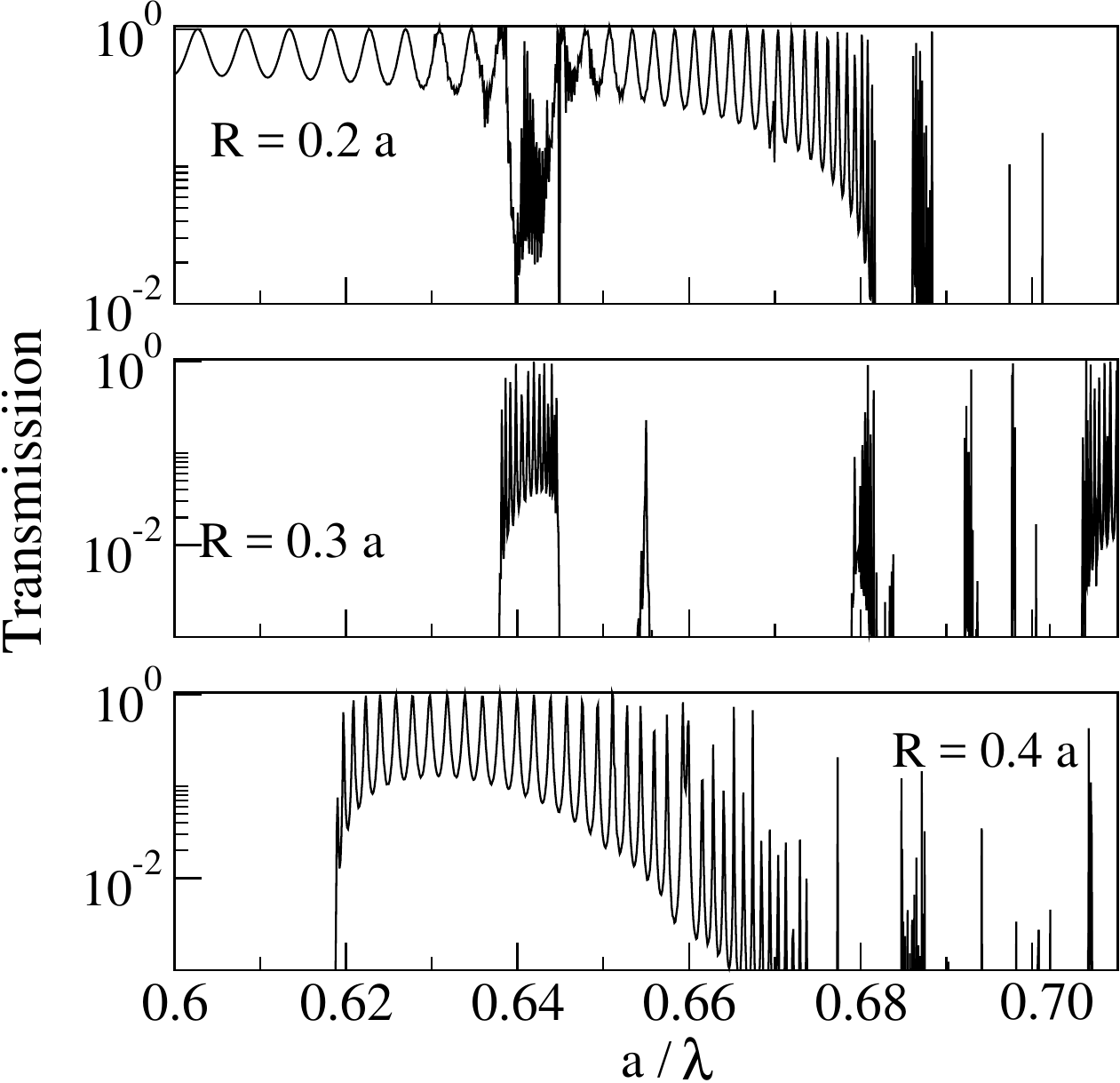}
~~
\includegraphics[width=0.23\textwidth]{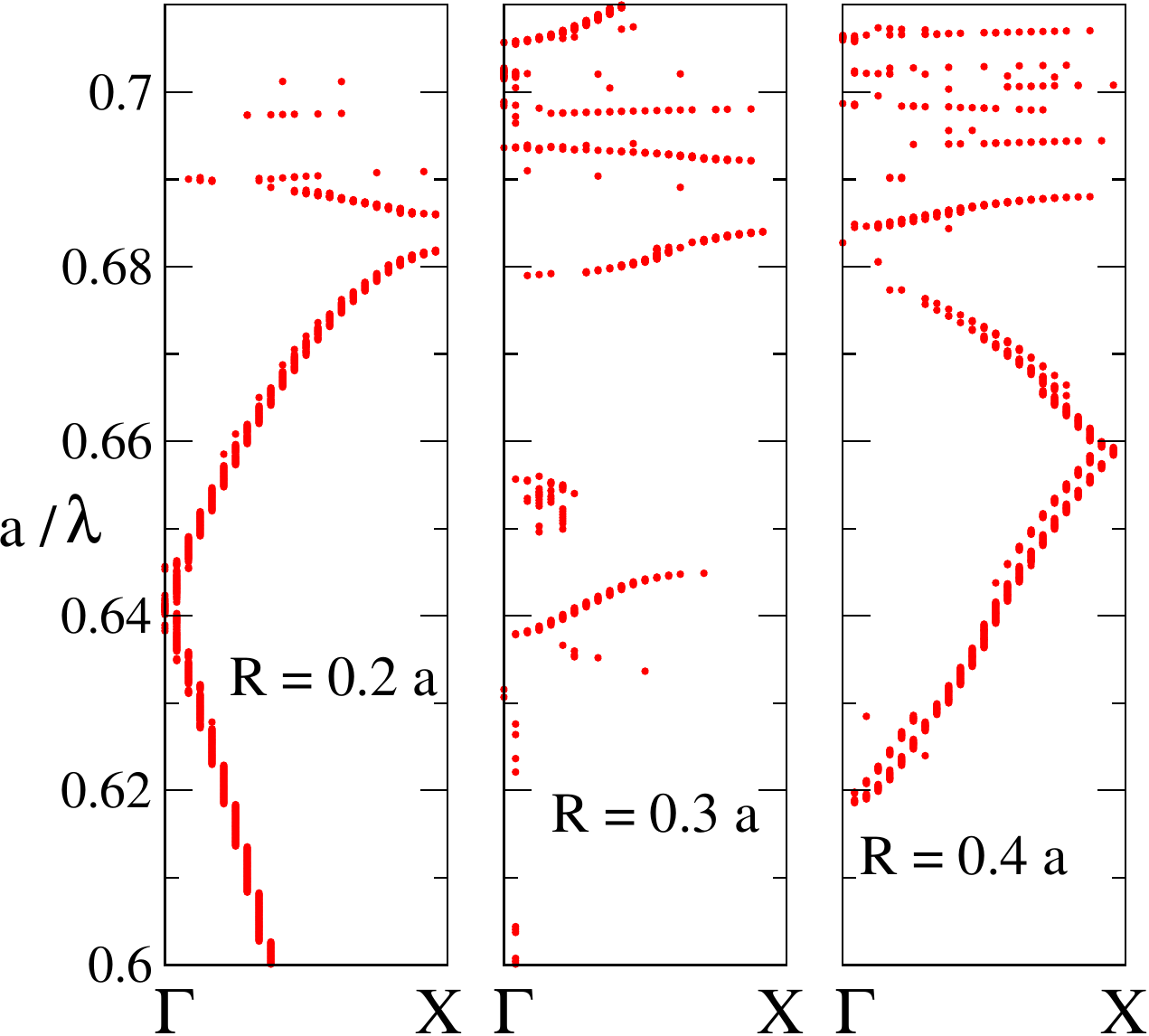}
\caption{Detail of Fig.~\ref{Hr_02_03_04}: The transmission coefficient (left) and band structure
for $N=24$ rows of cylinders in the frequency range $0.60<f<0.71$.
A group of
modes observed for $R=0.3a$ at frequency interval  $a/\lambda \approx  0.65$
 exhibits incomplete band which 
corresponds to surface waves will be explained in Sect. \ref{surface}.
}
\label{Hr_02_03_04-d1}
\end{figure}

\begin{figure}[t]
\noindent\includegraphics[width=0.23\textwidth]{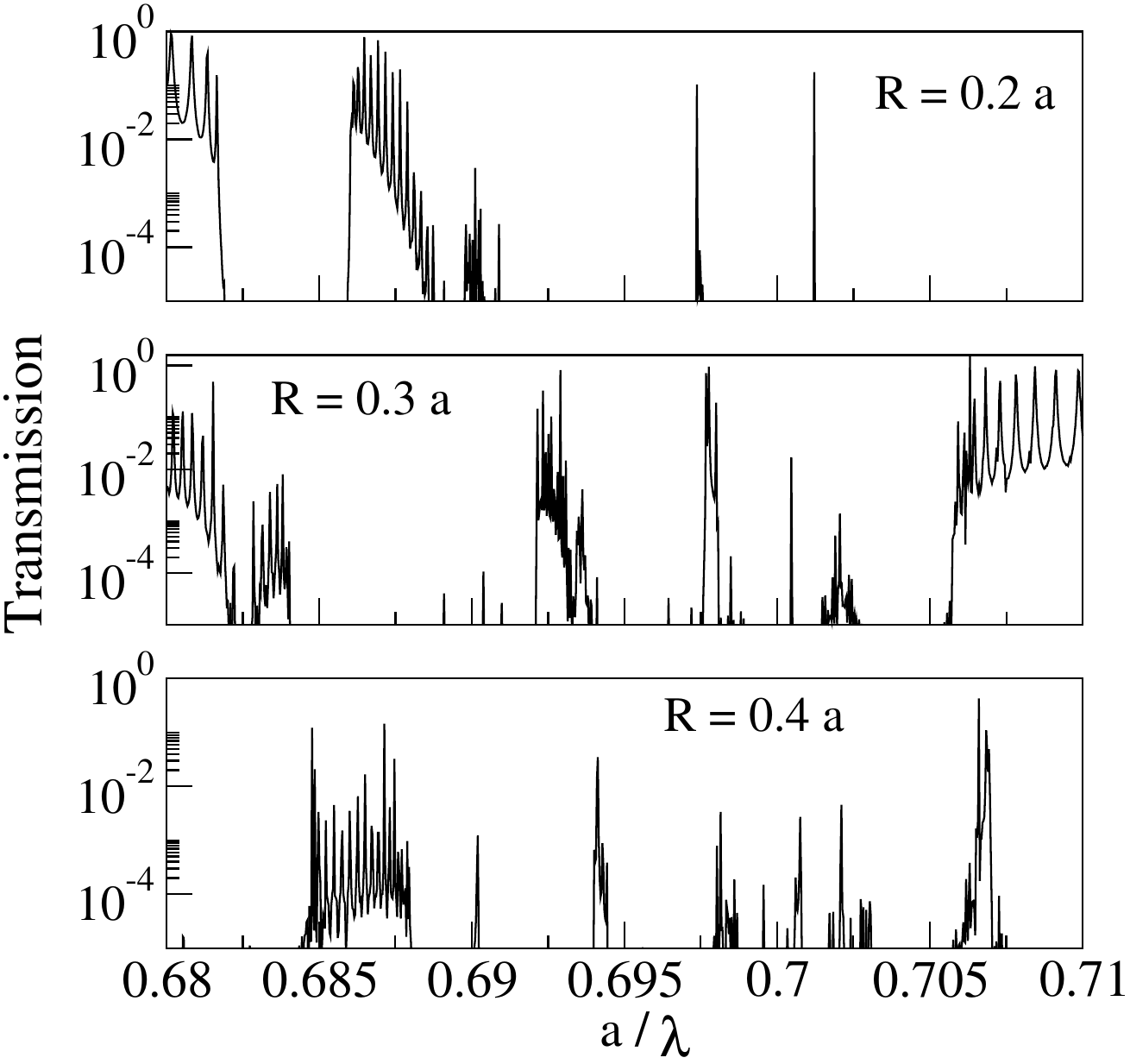}
~~
\noindent\includegraphics[width=0.23\textwidth]{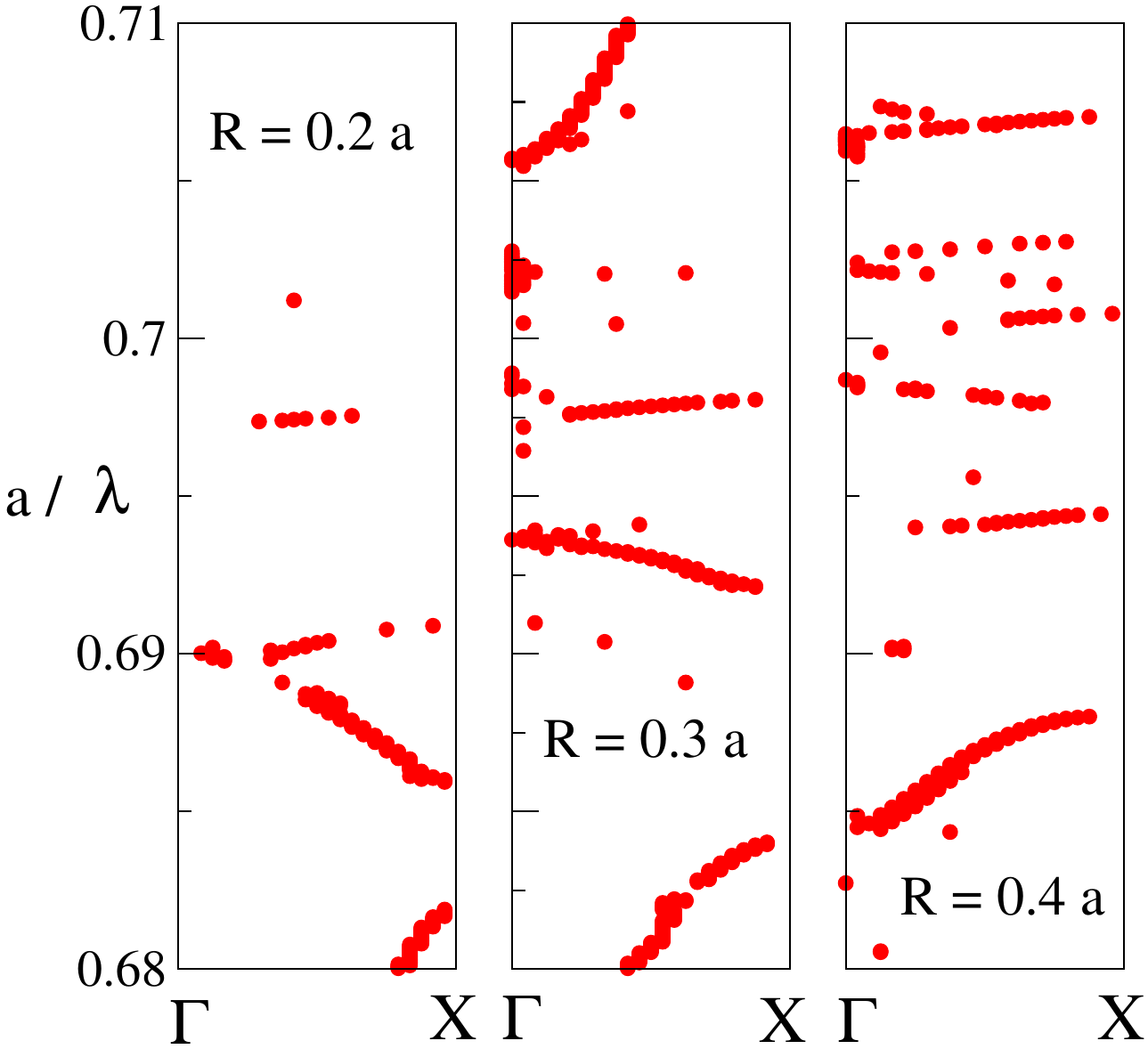}
\caption{Detail of Fig.~\ref{Hr_02_03_04-d1}: The transmission coefficient (left) and band structure
for $N=24$ rows of cylinders in the frequency range $0.68<f<0.71$.}
\label{Hr_02_03_04-d2}
\end{figure}

\subsection{Band structure}

The results depicted in Fig.~\ref{H-cylinders} demonstrate that by increasing the radius of cylinders
the frequencies of the lowest Mie resonance decreases and thus provides a wider frequency range where plane-wave-like bands interact
with Fano bands.  This trend can be also tracked down in the behavior of the band structures and transmittances
for all three values of the radius shown in Fig.~\ref{Hr_02_03_04} within the frequency range $0 < f < \omega_p$ and in Figs.~
\ref{Hr_02_03_04-d1} and \ref{Hr_02_03_04-d2} which zoom-in at the frequency range $0.68 < f < 0.71$ , where the coupling between
Bragg and Fano bands takes place.

When $R = 0.2a$,  the band structure mostly consists of four broad frequency bands. The lowest one, which appears in the frequency range $0 < f < 0.37$
is undoubtedly the first Bragg band \cite{fP}. The spectrum in the frequency range above the first band gap contains the Bragg band which starts from the $X$-point
in the first Brillouin zone at the frequency $f = 0.49$. In fact, this band is split into two parts: the lower one within the frequency range $0.49 < f < 0.64$
and the upper one for the frequencies $f > 0.81$ -- see Fig.~\ref{Hr_02_03_04}. Above the former band we observe a broad Fano band in the frequency range
$0.64 < f < 0.68$ which corresponds to the 2nd resonance associated with a cylinder  - see Fig.~\ref{H-cylinders} and a series of narrow higher order Fano bands.
Due their narrowness, only a few of them have been identified numerically. Some of them are shown in Figs.~\ref{Hr_02_03_04-d1}  and
\ref{Hr_02_03_04-d2} which show details of the band structure.
The presence of the broad Fano band is resembled in the transmittances which display a strong interference pattern in the corresponding
frequency range shown in left panels in Figs.~\ref{Hr_02_03_04} and \ref{Hr_02_03_04-d1}.
The conjecture based on the splitting of the Bragg band described above is supported by the calculation of the band structure of periodic array
of dielectric rods with frequency-dependent permittivity $\varepsilon=a^2/\lambda^2-1$, which equals exactly to the opposite
value of metallic permittivity given by Eq.~\ref{Drude}. Indeed, the spectrum consists of two Bragg bands separated by gap at the frequency $f\sim 0.47$
\cite{markos-lviv}.
In the frequency region $0.68 < f < 0.707$ we expect a series of localized flat bands. Since 
the bands are very narrow, only a few of them is visible in Figs. 3--5.
(Note that the width of bands decreases when cylinder radius decreases.)
We have not studied them in detail, instead we focused on the band structure corresponding to the photonic crystal consisting of
the cylinders with radius $R=0.3a$.

\begin{figure}[h!]
\includegraphics[width=0.95\linewidth]{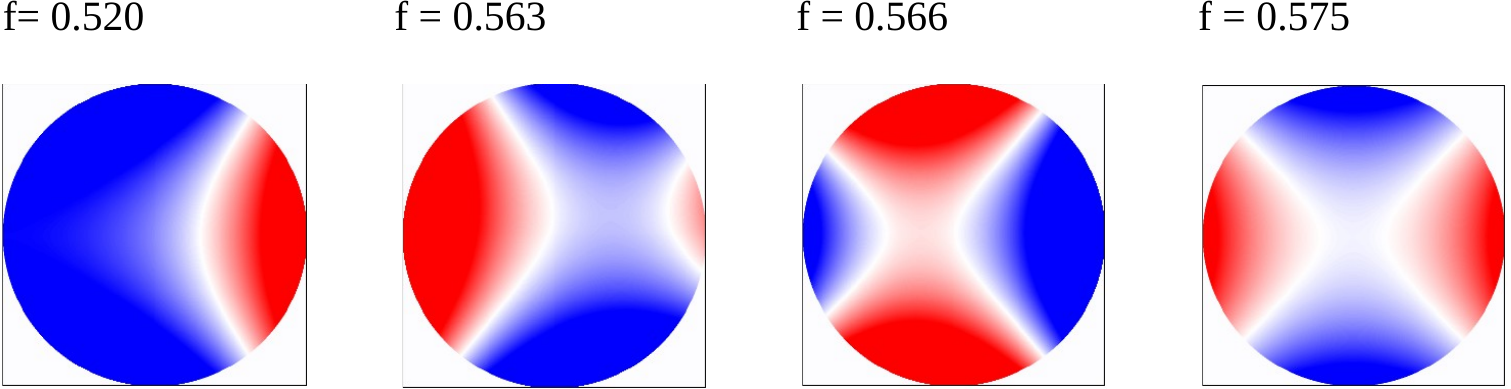}
\caption{(Color online) Field distribution in the 2nd band of $R=0.3 a$ photonic slab for frequencies
f = 0.520, 0.563, 0.566 and 0.575. The symmetry of field changes when frequency increases from the lower band edge $f = 0.47$ to upper one
$f \simeq 0.578$.
Here and in all subsequent Figures blue (dark gray) and red (gray) corresponds to positive and negative values of $H_z$, respectively and the
EM wave propagates from the left to the right.
}
\label{fig-2nd}
\end{figure}

The results shown in the right panel Fig.~\ref{Hr_02_03_04} show the band structure is significantly modified when radius of the cylinder is
increased: the two lower bands occur in the domain where Bragg-like scattering constitutes a dominant scattering mechanism
and a structural gap between the two lowest bands becomes wider due to the larger value of the filling fraction.
Likewise in the previous case for $R = 0.2a$,  we observe two separated parts of the second lowest Bragg band arising from splitting:
the lower one within the frequency range $0.5 < f < 0.59$ and the upper one for the frequencies $f > 0.93$ -- see Fig.~\ref{Hr_02_03_04}.
We note that the field distribution associated with the band within the frequency range $0.47 < f < 0.578$ at the $X$-point mostly reflects
behavior of the $A_1$ extended mode -- see Fig.~\ref{fig-2nd}. When the frequency is increased, the field distribution gradually transforms
into the pattern with a stronger angular dependence and can be identified as a modified plasmon resonance with the angular dependence $\cos 2\theta$
-- see the magnetic fields shown in Fig.~\ref{fig-2nd}.
The scale used in field distribution plots in Fig.~\ref{fig-2nd} as well as in all other field distribution snapshots hereafter in the paper
is not uniform and is given in arbitrary units. For instance, the field in left panel in Fig.~\ref{fig-2nd} varies between $-$1 and $+1$, while
the maximal amplitude of field in the right panel is only 0.1. We note, that the difference in the amplitude of the fields shown in Fig.~\ref{fig-2nd}
resembles the varying character of the mode, while the different scales in all other figures reflect symmetry of the bands and their coupling to an incident
plane wave.


The enhanced interaction in the frequency range $0.6 < f < 0.707$ leads to a strong fragmentation band structure in this frequency range.
Specifically, when $R = 0.3a$ we observe a  number of very narrow localized modes which belong to both symmetric and antisymmetric modes listed in Table I which
can be assigned to the irreducible representations of $C_{4v}$  symmetry group. The transmittances shown on the left panels in
Figs.~\ref{Hr_02_03_04-d1} and \ref{Hr_02_03_04-d2} reflect the existence of these localized bands. In the case of the largest radius considered
$R = 0.4a$ the localized states occur in wider frequency range starting from $f = 0.5$ and the spectrum consists of the
increasing number of localized modes converging to surface plasmon frequency $f = 0.707$ - see right panel in Fig.~\ref{Hr_02_03_04}

\begin{figure}[t]
\noindent\includegraphics[width=0.23\textwidth]{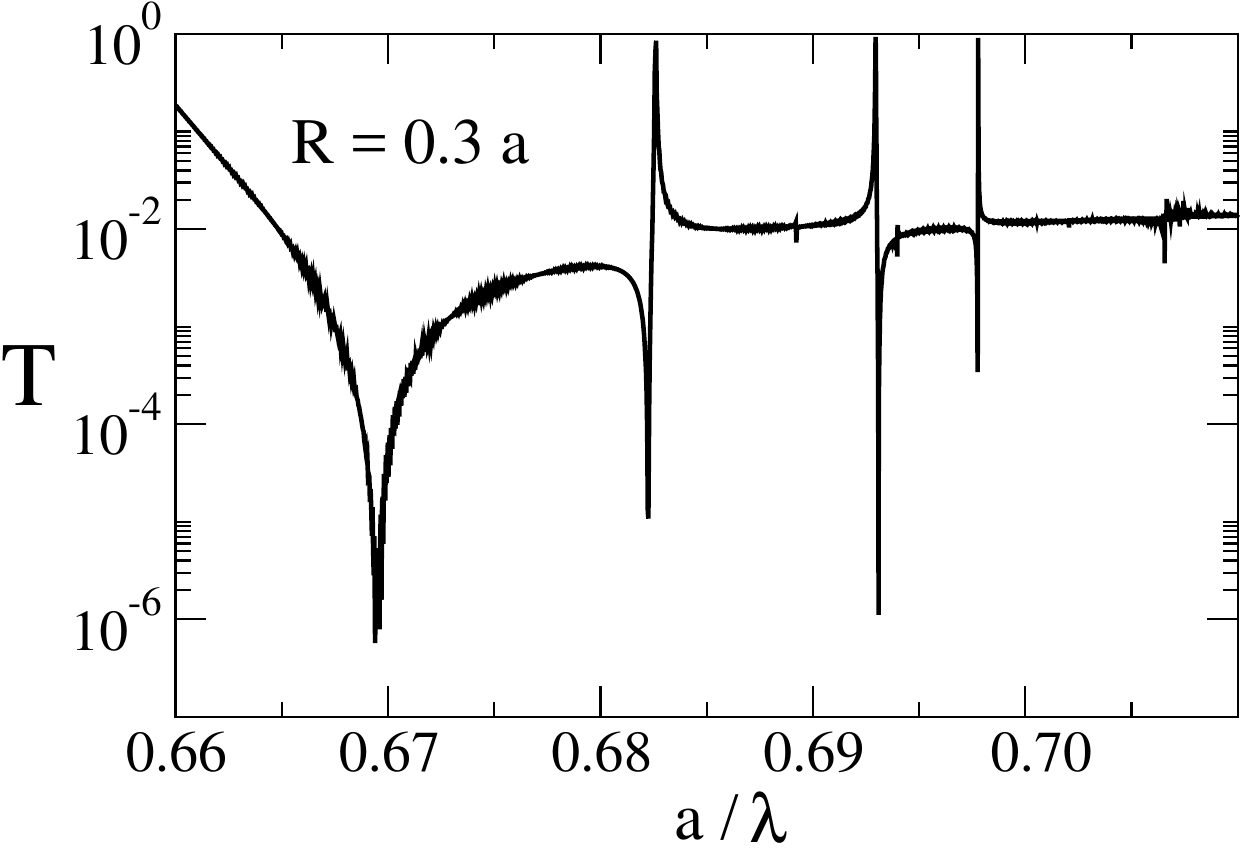}
\caption{The transmission of plane EM wave through a linear chain of metallic cylinders with radius $R = 0.3a$.}
\label{1riadok}
\end{figure}

\begin{figure}[t]
\noindent\includegraphics[width=0.19\textwidth]{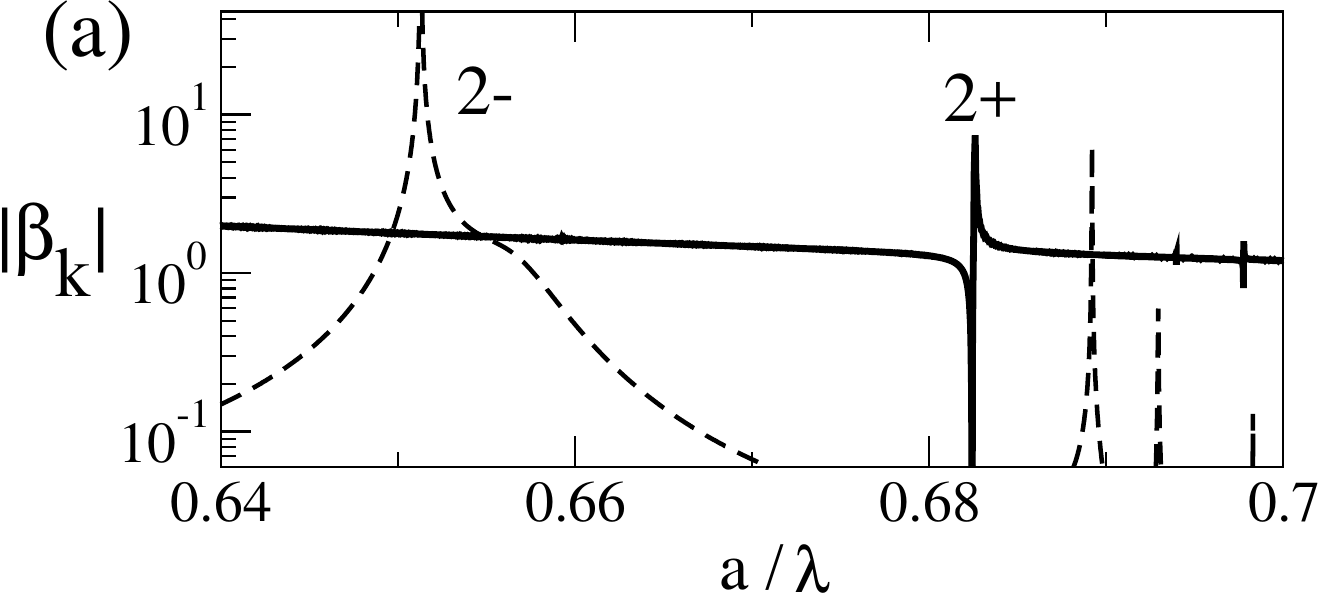}
~~~
\noindent\includegraphics[width=0.19\textwidth]{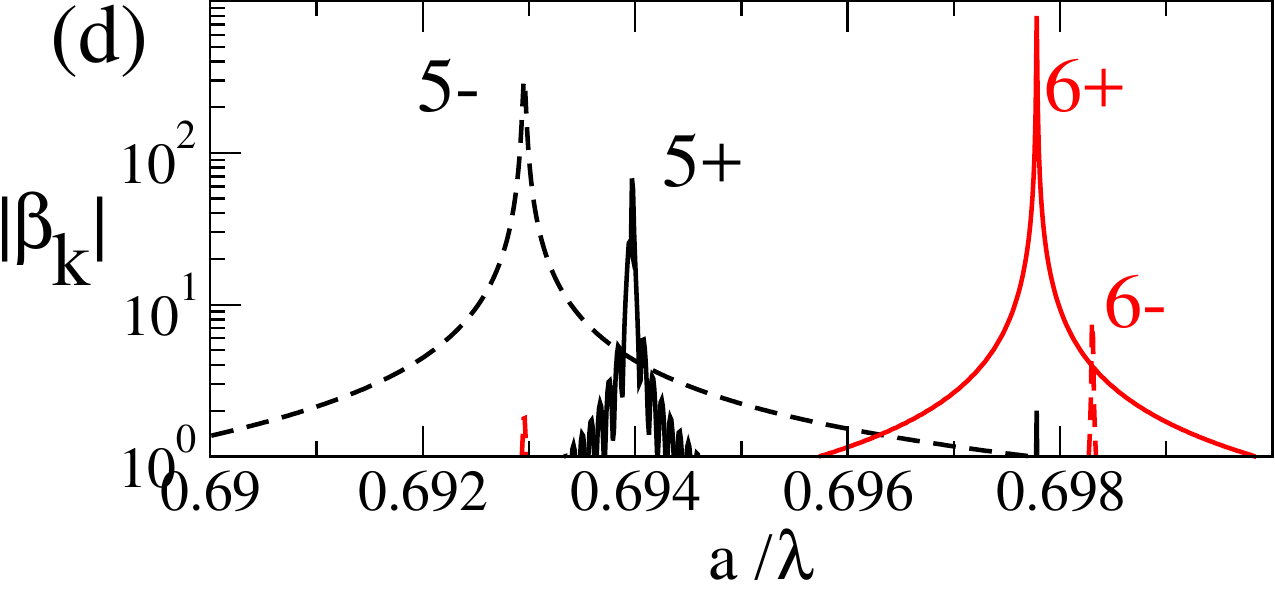}\\
\noindent\includegraphics[width=0.19\textwidth]{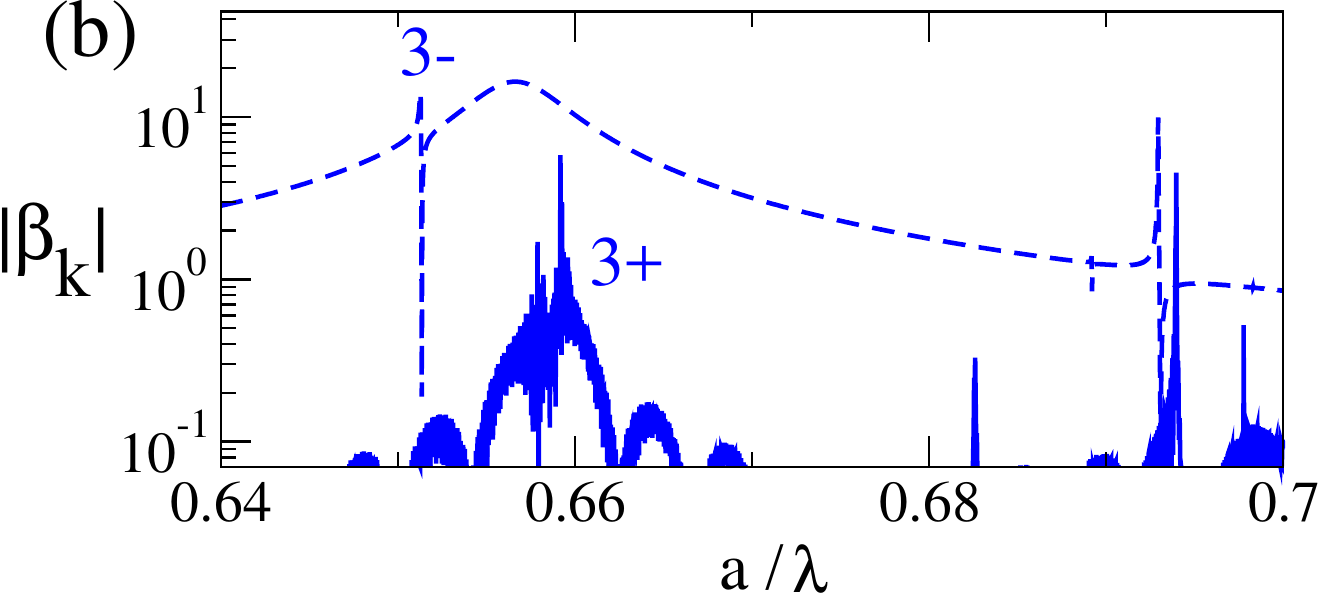}
~~~
\noindent\includegraphics[width=0.19\textwidth]{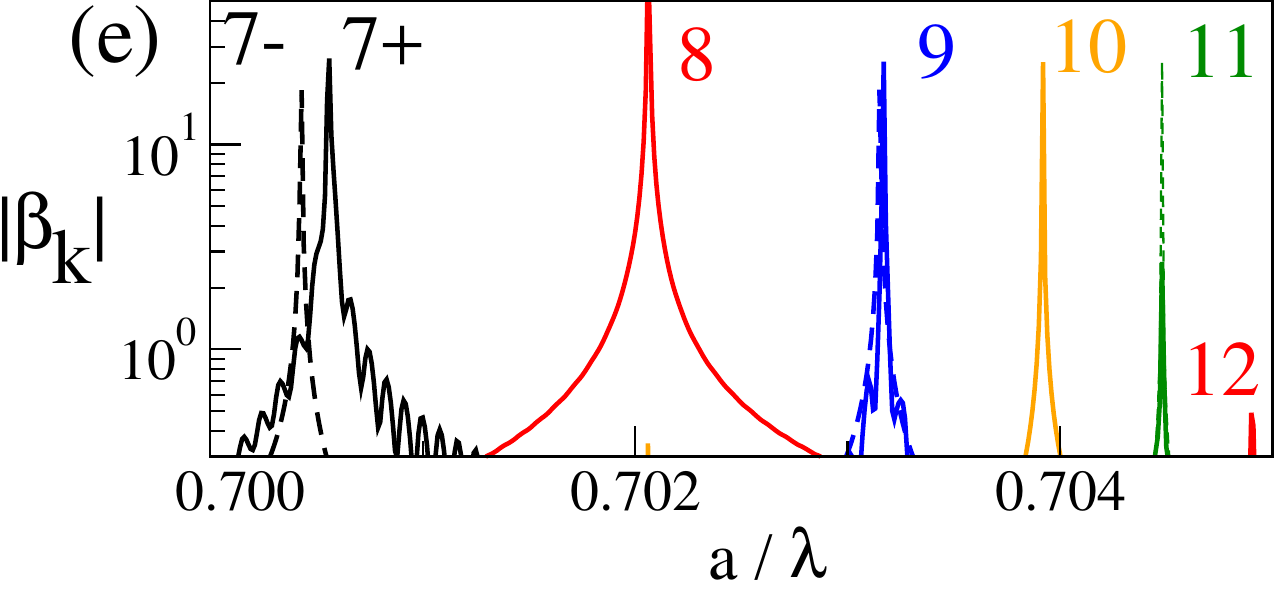}\\
\noindent\includegraphics[width=0.19\textwidth]{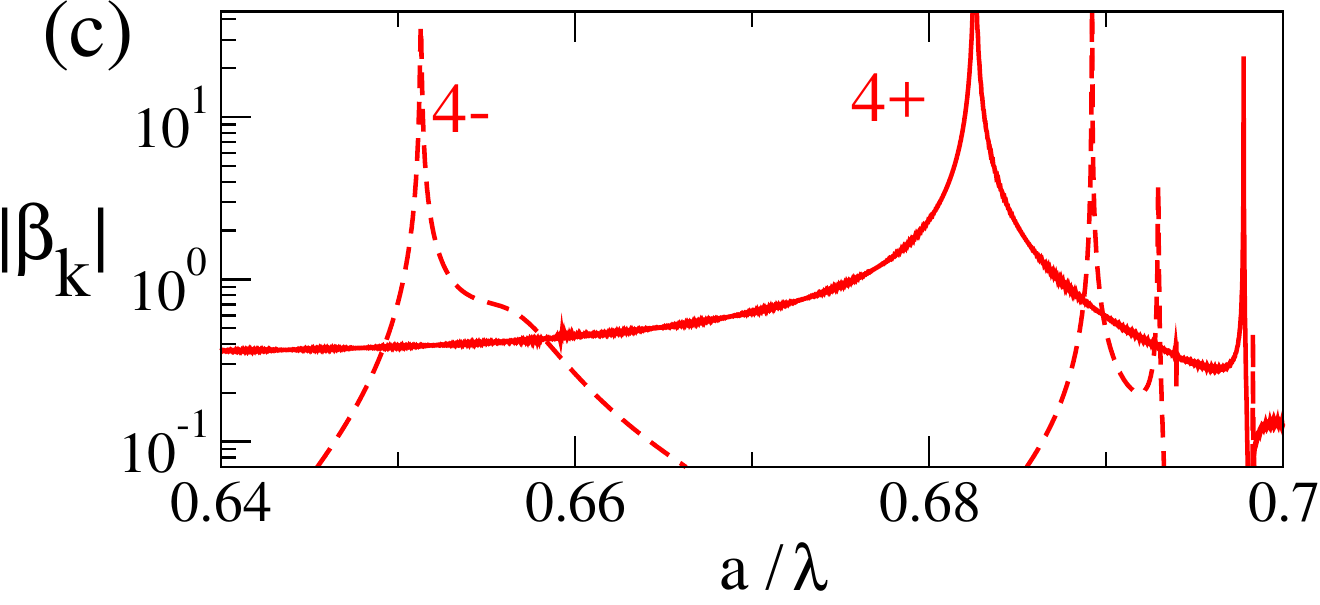}
~~~~~~~~~~~~~~~~~~~~~~~~~~~~~~~~~~~
%
\caption{Resonances excited at the linear chain of cylinders of radius $R=0.3a$.
Shown are coefficient $\beta^+$ (solid lines) and $\beta^-$ (dashed lines) calculated from the simulation of the propagation of the plane wave.
Note that some lower resonances overlap.
For instance, lower resonance 4- (panel (c)) overlaps with the resonance 2- shown in panel (a). The splitting of resonances to even and odd is clearly visible for $n=5,6$ and 7.
For higher resonances, the splitting is very narrow and two resonant peaks are nearly identical.
Some odd resonances possess small magnitude due to a small incident angle
$\theta=\pi/100$.}
\label{koeficienty}
\end{figure}

\subsection{Symmetry and formation of Fano bands}

To explore mechanisms underlying formation of Fano bands we first examine correspondence between Fano resonances in the transmittance for a linear array of metallic cylinders  and $\beta^{\pm}$ coefficients in the expansion of the scattered field given by Eq. \ref{eq.1b}. The transmittance for a linear chain of metallic cylinders with radius $R = 0.3a$ shown in Fig.~\ref{1riadok} reveals several Fano resonances that can be assigned to sharp peaks of the coefficients $\beta^{\pm}$ that are shown in Fig.~\ref{koeficienty}  and listed in Table \ref{table1}. The resonances in the transmittance arise from the interference of the incident EM plane wave and the LCAO states consisting of the localized surface plasmon modes associated with an isolated cylinders\cite{note4}. At the same time, each of the Mie resonances shown for the case $R = 0.3a$ in Fig.~\ref{H-cylinders} can be assigned to a symmetric or asymmetric LCAO state as it is summarized in in Table \ref{table1}. One can expect that the frequencies of a symmetric and asymmetric LCAO states associated with a linear chain are smaller and larger than the frequency of the corresponding Mie resonance of an isolated cylinder, respectively. This applies to all orders $n$ listed in Table \ref{table1} except that for $n = 3$ which corresponds to a non-LCAO state confined at the opposite sides of the photonic structure as we demonstrate below. To this end we also note that each state is assigned either to even or odd resonance depending on its order $n$ as we discuss below.

For example, one can assign the Mie resonance $n = 2$ at $f  = 0.61512$  to the symmetric state identified by resonance of $\beta_2^+$ at $f = 0.579$ and to the antisymmetric state identified by resonance of $\beta_2^-$  at $f  = 0.65136$.
The results for $\beta^{\pm}$ resonances associated with a linear chain of cylinders shown in Fig.~\ref{koeficienty}(a) reveal that for lower frequencies in the range $0.63 < f  < 0.69$ the coefficients display a significant overlapping. In addition, the maxima corresponding to the coefficients $\beta^{\pm}$ of different orders may occur at the same frequency. Therefore, the Mie resonances listed in Table \ref{table1} are assigned to the resonance of $\beta^{\pm}$ coefficient corresponding to a leading term at a given frequency. It is important to keep in mind, that although the field pattern of the modes typically reflect a symmetry of the leading terms in the LCAO expansion, generally in the interpretation of resulting Fano bands one has take into account all terms, in particular in the frequency range where overlapping of the $\beta^{\pm}$ resonances occurs.

The symmetry of the eigenmodes of 2D band structure can be classified in terms of the resonant states associated with an isolated cylinder.  Likewise the eigenmodes associated with linear chain the eigenfunctions belonging to 2D band structure along $\Gamma-X$ direction in the first Brillouin zone are symmetric or asymmetric along $x$ axis. Simultaneously, the angular dependence of the resonant states which correspond to the surface plasmon modes which carry no momentum in the lengthwise direction of the cylinder\cite{Economou}, is described by $e^{\pm in \theta}$ for each $n \ge 1$  form symmetric and asymmetric combination corresponding to the functions $\sin n\theta$ and $\cos n\theta$, respectively. We note that $n$ defines symmetric functions $\sin n\theta$ and $\cos n\theta$, when $n$ is odd and even, respectively, while $\sin n\theta$ and $\cos n\theta$ are asymmetric when $n$ is even and odd, respectively. The different localized resonant modes labeled with the same irreducible representation can be distinguished in terms of number of lobes $n_L$  which define couples given by the functions $\sin \frac{n_L}{2} \theta$ and $\cos \frac{n_L}{2} \theta$, where $n_L = 2n; n \in N$ \cite{EM}.
The degeneracy of the eigenvalues corresponding to modes with $n_L = 4k$  of the isolated cylinder is lifted when they are arranged in the square lattice.while those corresponding to $n_L \ne 4n$  will have a twofold degeneracy.
For example, LCAO combinations of Mie resonant states with $n = 6$ form symmetric and asymmetric singlets $B_1(12)$ and $B_2(12)$ \cite{EM} in the square lattice as number of lobes $n_L = 4n$ -- see Fig.~\ref{beta-6}, while LCAO combinations of Mie resonant states with $n = 5$ form two-fold degenerate  state $E(10)$ since number of lobes  $n_L \neq  4n$. The degeneracy of the $E(10)$ state is lifted along $\Delta$ direction and the $E(10)$ state splits into a symmetric and an asymmetric $A_1(10)$ and $B_2(10)$ modes \cite{EM} -- see Fig.~\ref{beta-5}.

\subsection{2nd Fano bands}

\begin{figure}[t]
\noindent\includegraphics[width=0.23\textwidth]{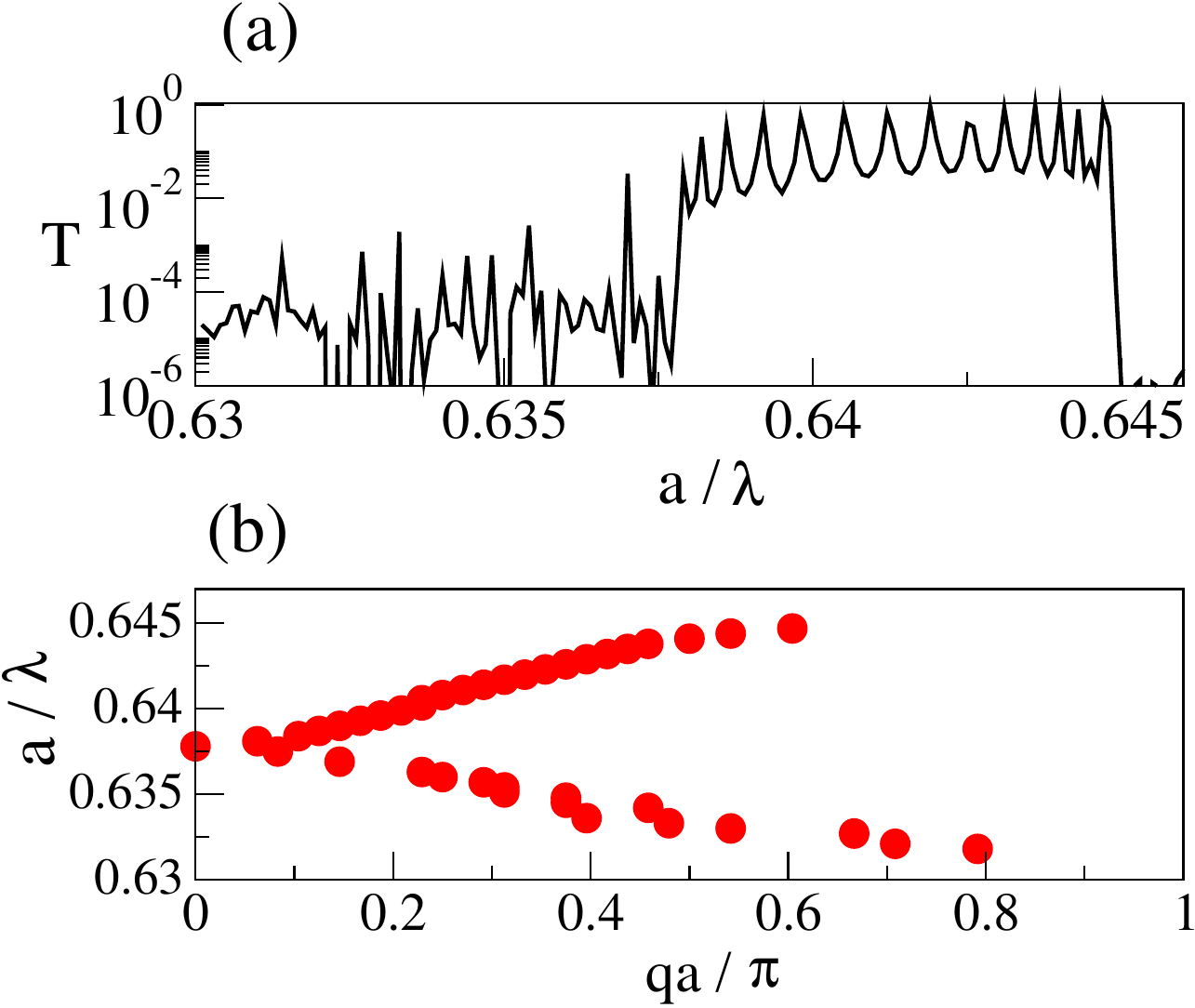}\\[3mm]
          \includegraphics[width=0.95\linewidth]{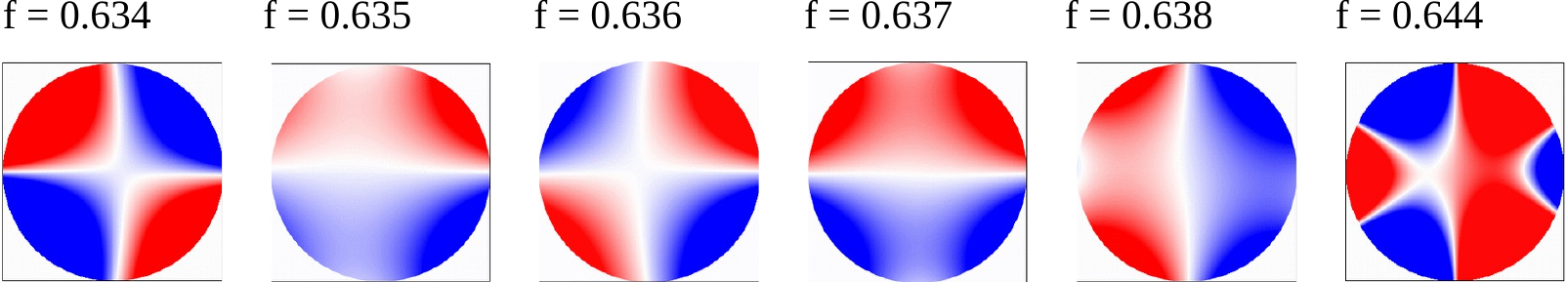}
\caption{(a) The transmission coefficient and two corresponding frequency bands
of symmetry $n = 2$ and $R=0.3a$. The transmission in the lower band is small due to its odd symmetry. (b) Two frequency bands recovered from transmission data. Lower panel shows the
field distribution at some frequencies inside the lower and upper band. 
The meaning of colors is given in caption to  Fig. 6.
}
\label{beta-2}
\end{figure}

Now we focus on the frequency range $0.63 < f  < 0.645$ in which a doubly degenerate $E(6)$
state at the $\Gamma$-point splits into an asymmetric $A_2(6)$ and a symmetric $B_1(6)$ state along the $\Gamma-X$ direction in the first Brillouin zone. A small amplitude of the transmittance of the 2D structure in the frequency range $0.63 < f  < 0.6375$ shown in Fig.~\ref{beta-2} confirms an antisymmetric character
of the lower band with $A_2(6)$ symmetry while the large oscillations in the transmittance in the frequency range corresponding to symmetric band with $B_1(6)$ symmetry demonstrates its coupling with an incident plane wave.
 The field distribution associated with $A_2(6)$  band along the $\Gamma-X$ direction in the frequency range $0.63 < f  < 0.645$  indicates symmetry corresponding to $\sin 2\theta$ and $\sin \theta$ - that resemble the asymmetric 2- and 1+ states, respectively, while the higher band $B_1(6)$  in the frequency range $0.638 < f  < 0.645$
reveals the symmetry $\cos \theta$ and $\cos 3\theta$  that resemble the symmetric 1- and 3- states, respectively.

\subsection{Surface waves and transmission}\label{surface}

\begin{figure}[t]
\includegraphics[width=0.18\textwidth]{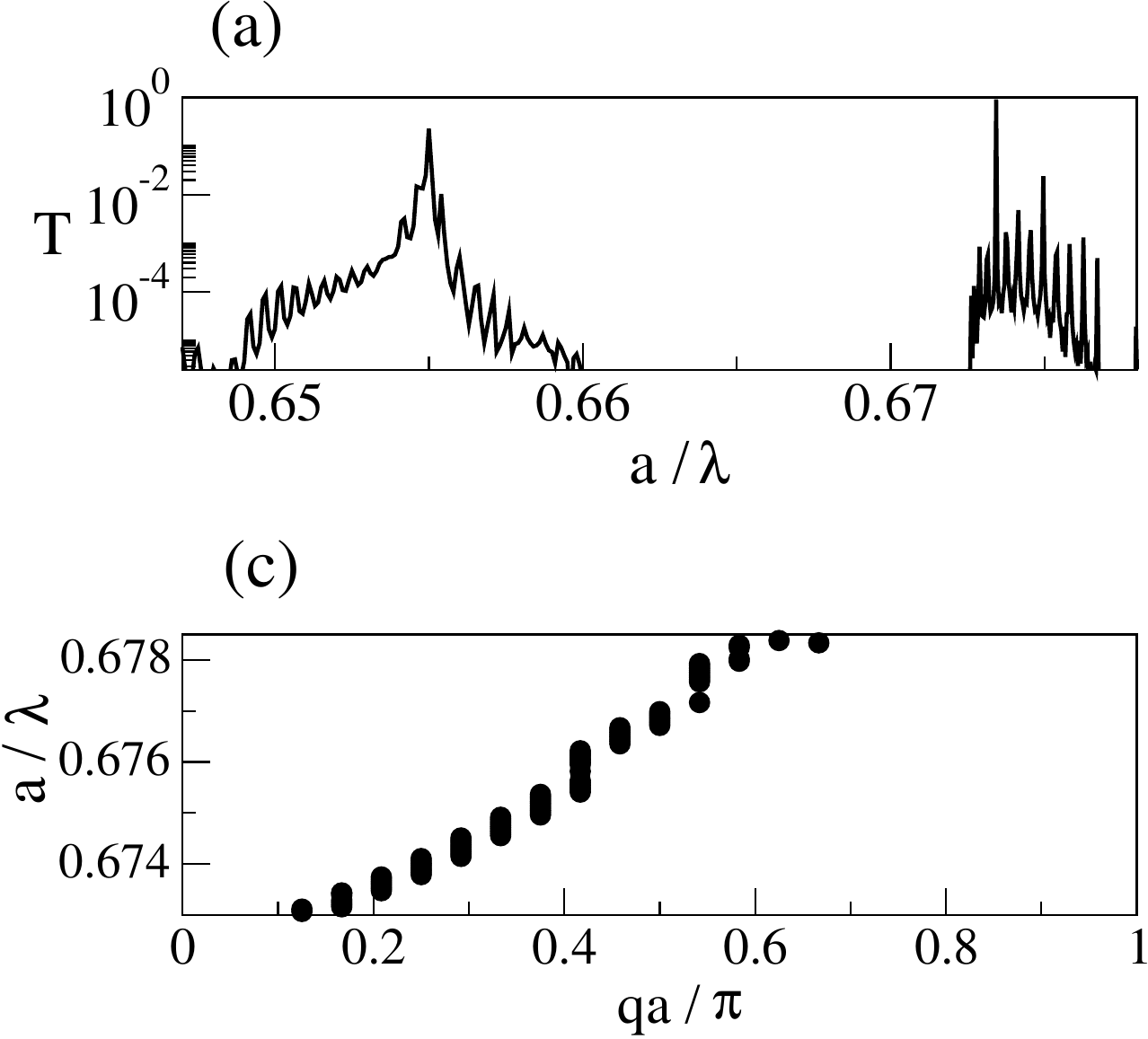}
~~
\includegraphics[width=0.3\linewidth]{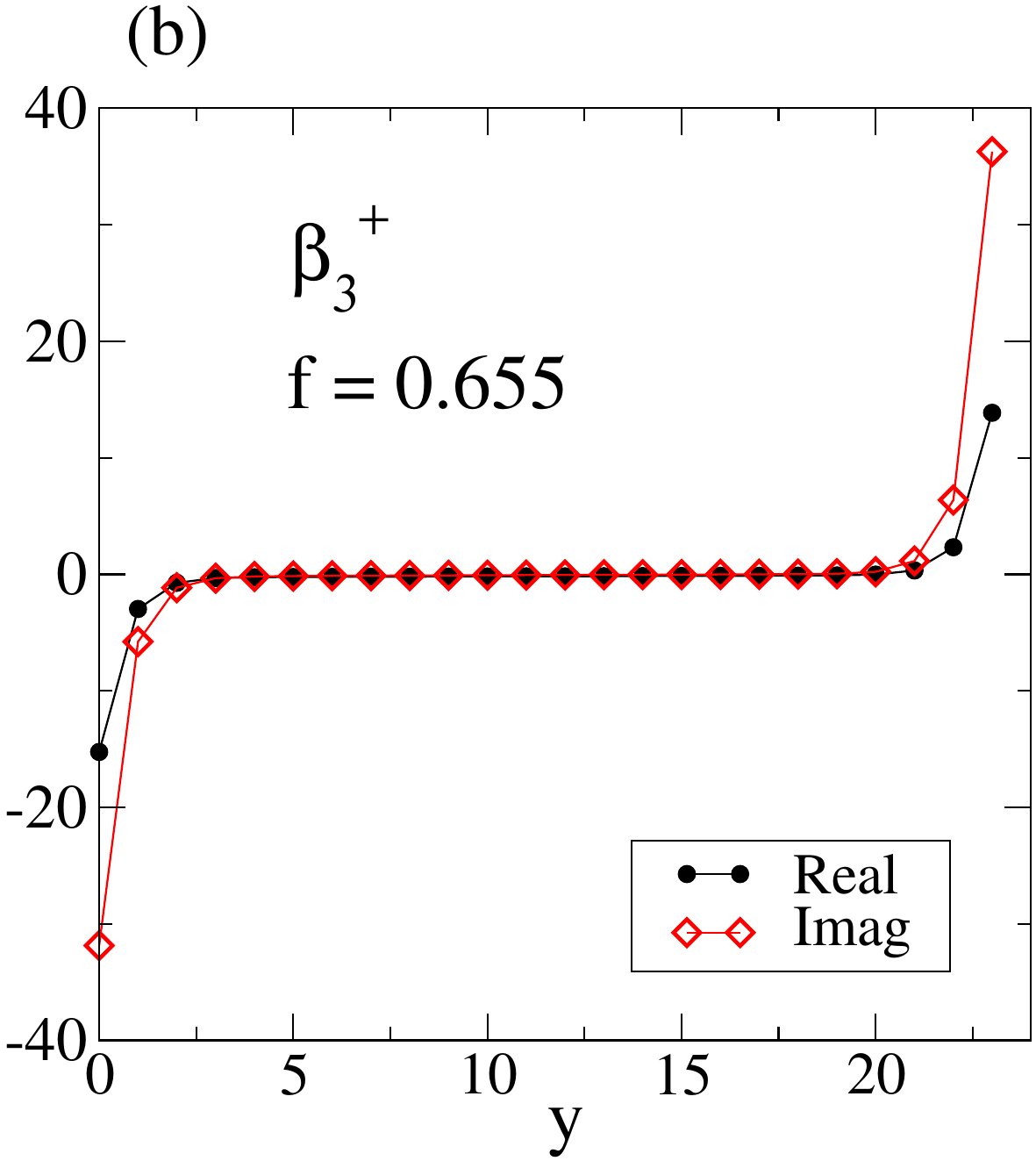}
\caption{(a) The transmission coefficient and coefficient $\beta_3^+$  in the region of the 3rd resonance.
$N=24$. (b) The spatial distribution of coefficient $\beta_3^+$ across the sample confirms that the transmission in the left transmission band
is due to the excitation of surface waves at the boundary of the entire photonic structure. Therefore, the band around the frequency $f=0.65$ cannot
be found by using eigenfrequency solver. (c) Frequency band around the frequency $f=0.676$.
}
\label{beta-3p}
\end{figure}

\begin{figure}[t]
\includegraphics[width=0.80\linewidth]{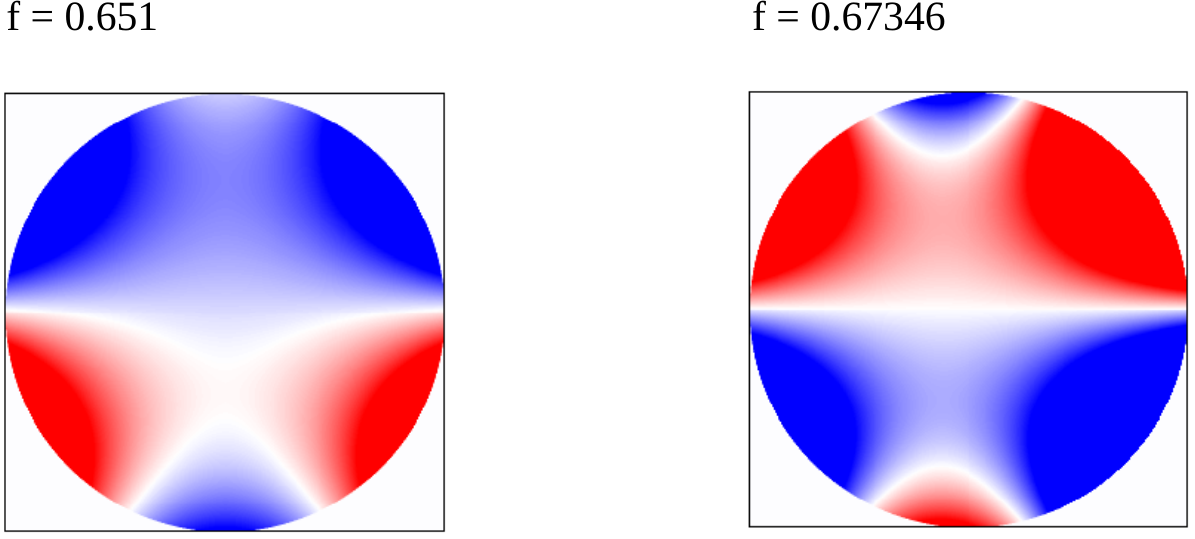}
\caption{(Color online)
The distribution of the
magnetic field $H$ at the frequency $f=0.651$  and $f = 0.67346$.
The meaning of colors is given in caption to  Fig. 6.
Owing to odd symmetry of the field and small incident angle, $\theta=\pi/100$,  the absolute value of magnetic field is small,  $|H_z|\le 0.02$.
}
\label{beta-3px}
\end{figure}

Of particular interest is the transmittance for $R = 0.3a$ in the frequency range $0.65 < f < 0.66$ which reveals somewhat irregular behavior shown in the left panel
Fig.~\ref{Hr_02_03_04-d1} and Fig.~\ref{beta-3p}(a). It resembles neither typical Bragg nor Fano band and shows an incomplete band.
A detailed analysis of the transmission coefficient is presented in Fig.~\ref{beta-3p}(b).
The field distribution of the mode with frequency $f = 0.651$ 
shown in the left panel of Fig.~\ref{beta-3px} 
displays the symmetry which corresponds to the asymmetric
3+ state, however the amplitude of the field is very small.

To unveil the origin of such a peculiar behavior we exploited the spatial distribution of the $\beta_3^{+}$  resonance [Fig.~\ref{beta-3p}(b)] which displays enhancement of the amplitude at the boundaries of the structure while the amplitude inside the structure is vanishing. This suggests, that the eigenvalues in this frequency range correspond to the surface modes propagating along the opposite interfaces of the photonic structure. In addition, one can observe the two characteristic features: 1. the surface modes are coupled through the tunneling which occurs inside the structure 2. the field does not reveals oscillations due to an exponential decay of the field with increasing distance from the surface into the structure. As a result, these modes cannot be obtained from Eq. (\ref{eq.q}) and consequently do not span over entire first Brillouin zone as is shown in Fig.~\ref{Hr_02_03_04}.
It is interesting to note that these surface modes exist due to the finite size of our structure and our approach proved to be capable to deal with these modes which have not been reported in Refs. ~\onlinecite{KS} and ~\onlinecite{EM}.

The higher band indicated in the transmittance in the frequency range $0.673 < f  < 0.679$ shown in Fig.~\ref{beta-3p}(c) supports in contrast to the lower one, a localized mode. 
The field distribution  shown in the right  panel of Fig.~\ref{beta-3px} confirms that
it can be also assigned to asymmetric 3+ state and corresponds to the $B_2(4)$ reported in Refs.~\onlinecite{KS} and ~\onlinecite{EM}.

\subsection{The 4th bands}

\begin{figure}[t]
\begin{center}
\includegraphics[width=0.22\textwidth]{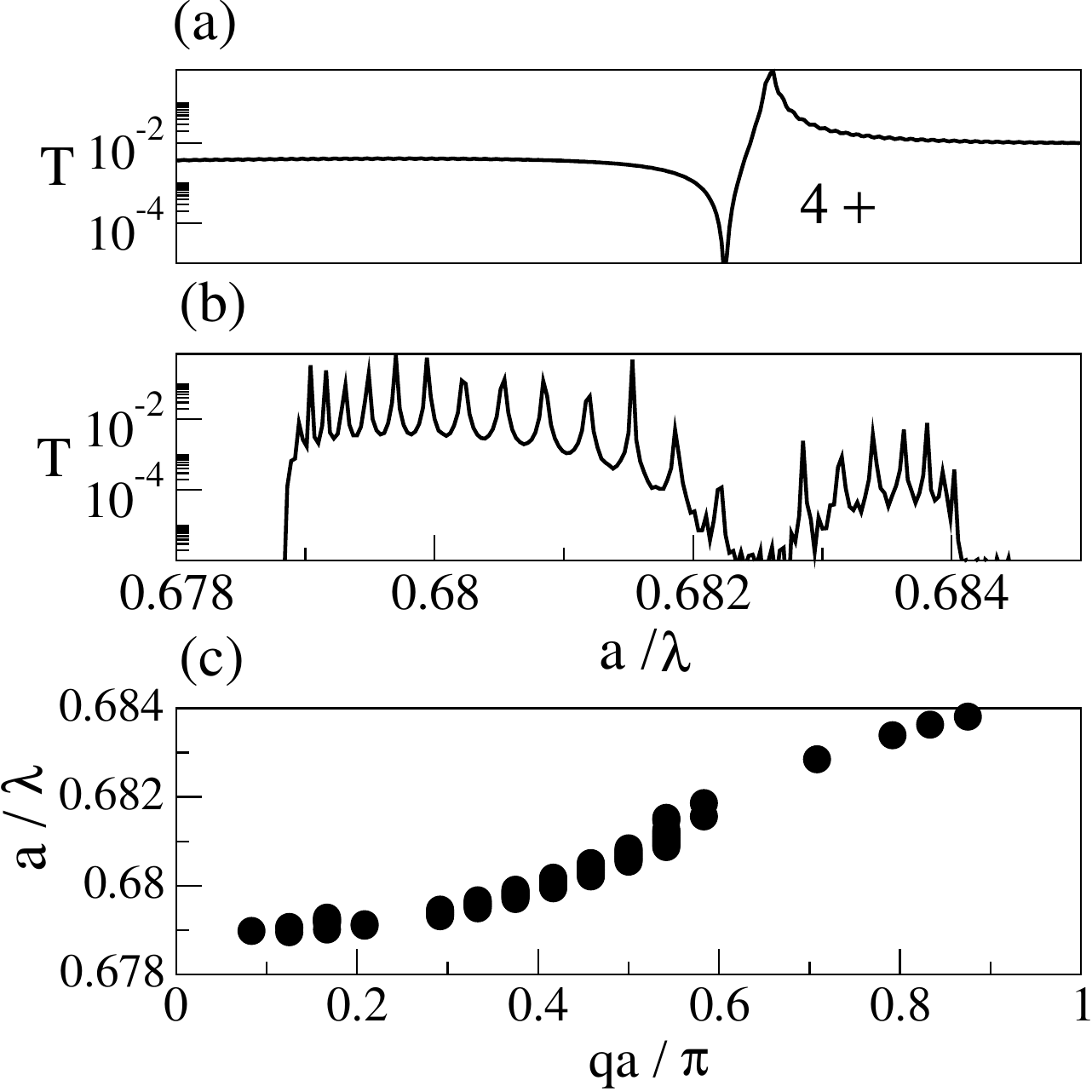}
~~
\includegraphics[width=0.24\textwidth]{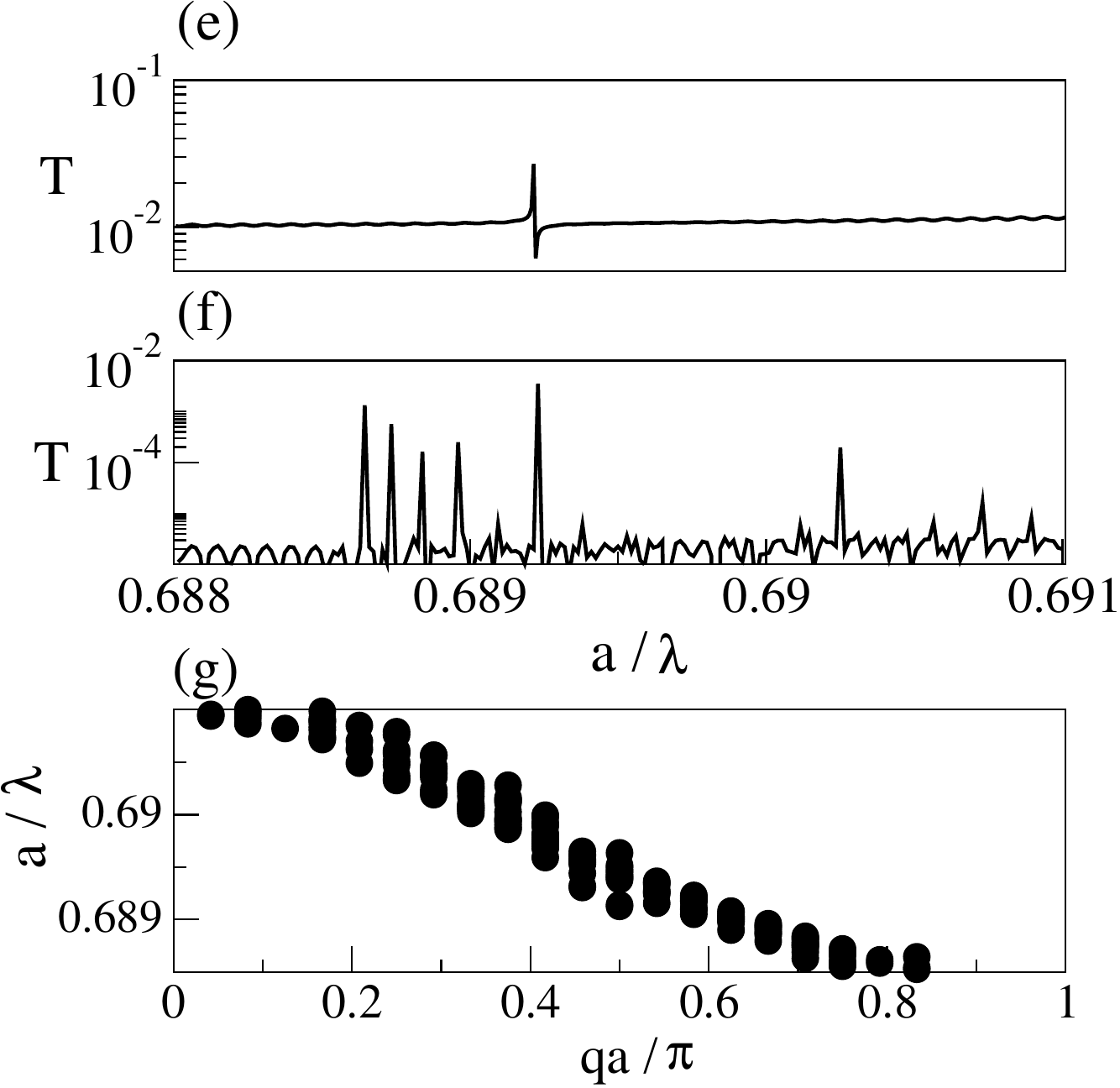}\\
\includegraphics[width=0.60\linewidth]{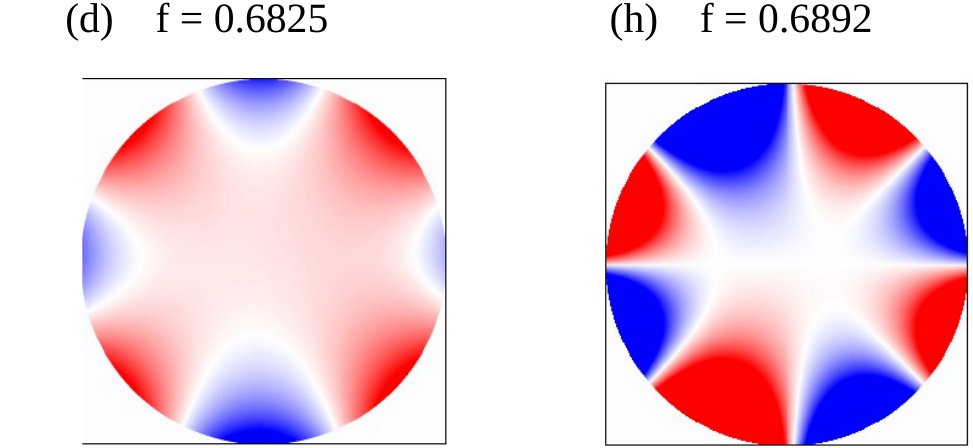}
\end{center}
\caption{(Color online) Left panel shows for the even band 4+ (a) the transmittance through linear array of rods, (b)~the transmittance through 24 rows,
(c)~the  frequency band and (d) The  magnetic field at the frequency $f=0.6825$. 
(The meaning of colors is given in caption to  Fig. 6.)
The right panel shows the same results in (e)-(g), except that for the odd
band 4-  and (h) magnetic field at the frequency $f=0.6892$.}
\label{beta-4}
\end{figure}

In Fig.~\ref{beta-4} we present the results which describe the band structure and transmittance in the frequency range $0.678 < f  < 0.684$. The lower band 4th band in this frequency range can be assigned to the symmetric 4+ state belonging to the resonance
of $\beta^{4+}$ coefficient shown in Fig.~\ref{koeficienty}(a) which is associated with Fano resonance in the transmittance of the linear chain shown in Fig.~\ref{beta-4}(a).
The transmittance of the 2D structure shown in Fig.~\ref{beta-4}(b) is significantly reduced in the upper part of the band within the frequency range $0.682 < f < 0.684$  and reveals a subgap which is correlated with an irregularity in the dispersion curve in the same frequency region shown in
Fig.~\ref{beta-4}(c). The field distribution displays $\cos 4\theta$ pattern -- see Fig.~\ref{beta-4}(d) -- which can be identified as a symmetric $A_1(8)$ state at the $\Gamma$-point reported in Ref.~\onlinecite{EM}.

The upper 4th band which we found in the frequency range $0.889 < f  < 0.691$ is shown in Fig.~\ref{beta-4}(g). It can be assigned to asymmetric 4- state belonging to the resonance of $\beta^{4-}$ coefficient shown in Fig.~\ref{koeficienty}(a) which is associated with Fano resonance in the transmittance of the linear chain shown in Fig.~\ref{beta-4}(e)
The vanishing transmittance of the 2D structure shown in Fig.~\ref{beta-4}(f) confirms the symmetry of the odd band 4-. The field distribution shown in Fig.~\ref{beta-4}(h) displays $\sin 4\theta$ pattern and which can be identified as asymmetric $A_2(8)$ state at the $\Gamma$-point reported in Ref.~\onlinecite{EM}.
We note that there is a large gap between these two bands which reflects a distance between the 4+ and 4- resonances -- see Fig.~\ref{koeficienty}(a).

\subsection{Higher order resonances}

\begin{figure}[h!]
\includegraphics[width=0.4\linewidth]{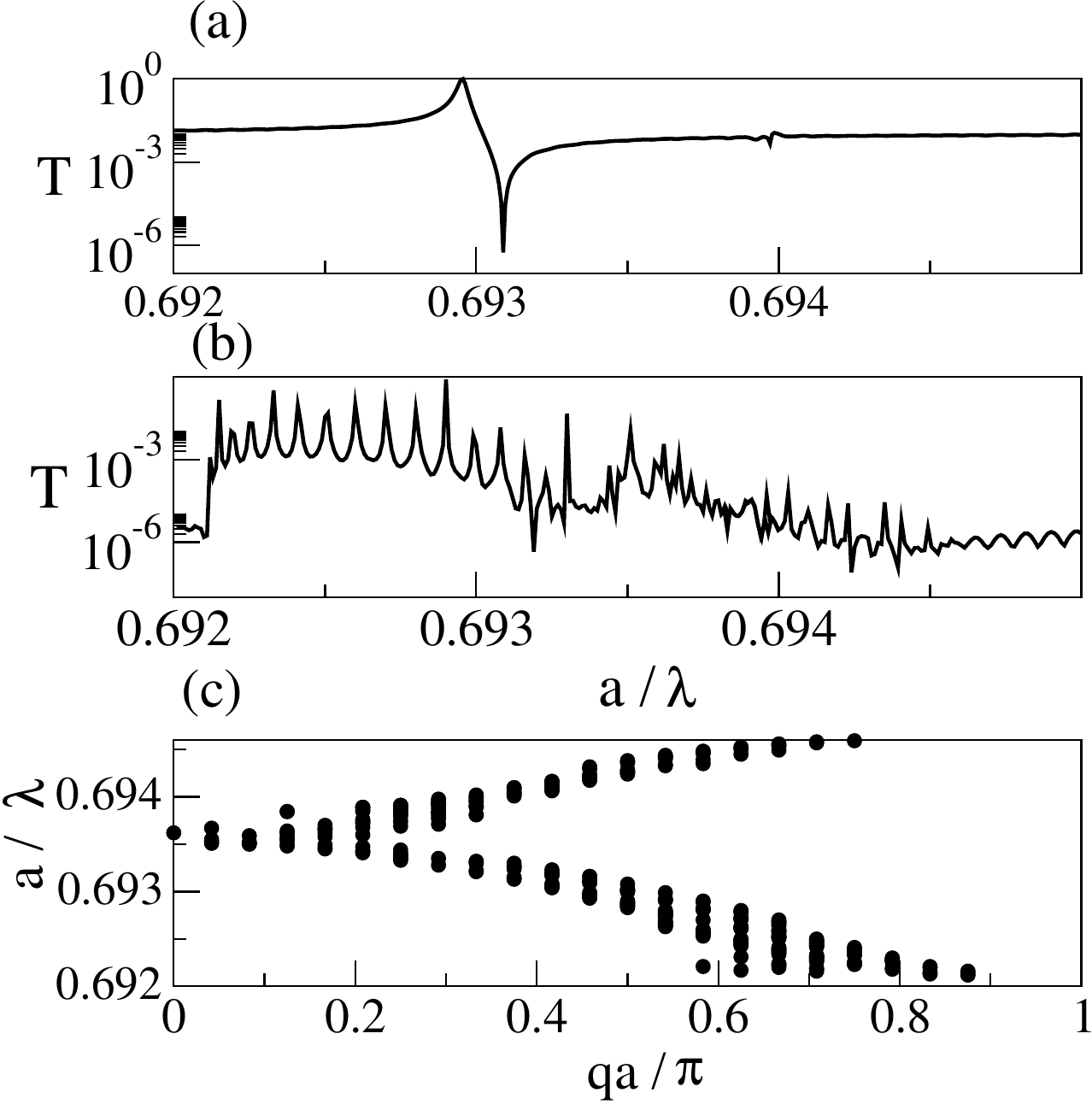}
~~
\includegraphics[width=0.45\linewidth]{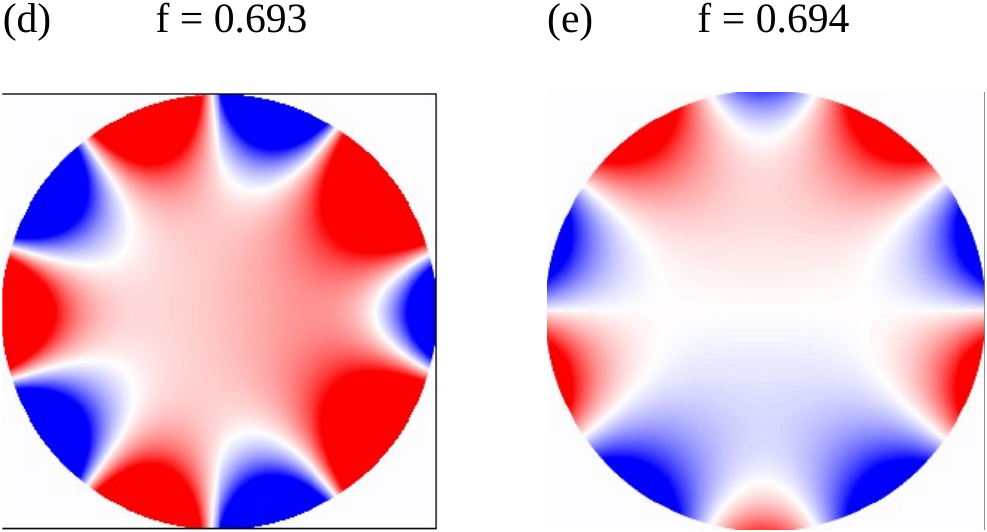}
\caption{(Color online) The 5th bands. Left panel shows (a)
the transmission of EM wave through linear array of metallic rods (note a tiny resonance at $f=0.694$), (b) the transmittance
through 24 rows of rods, (c) the symmetric and asymmetric frequency bands. The magnetic field at frequency
$ f = 0.693$ (lower symmetric band $5-$) and  at the frequency $f = 0.694$ (upper asymmetric band $5+$) is shown in right panels (d) and (e).
(The meaning of colors is given in caption to  Fig. 6.)
}
\label{beta-5}
\end{figure}

\begin{figure}[h!]
\includegraphics[width=0.4\linewidth]{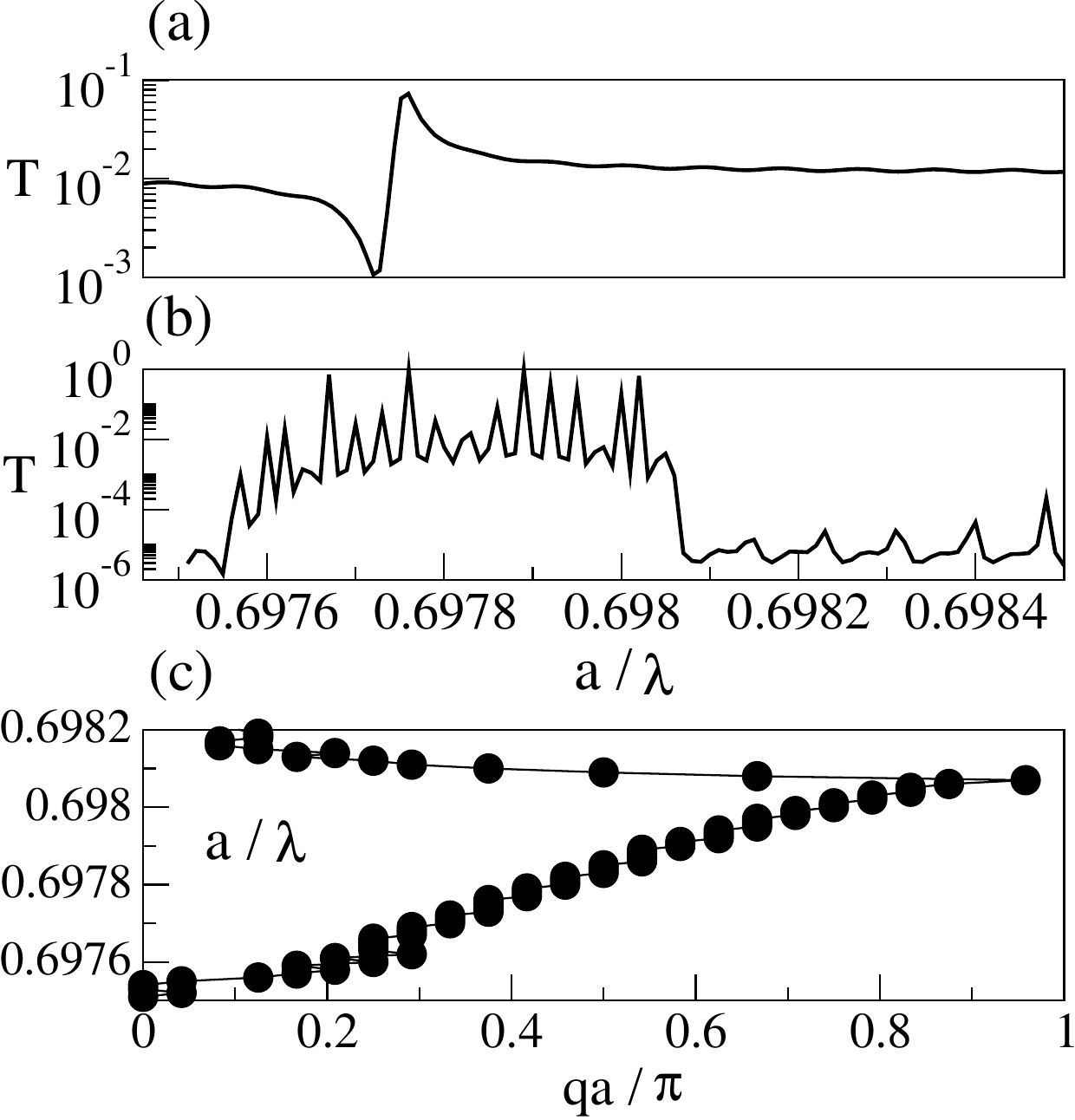}
~~
\includegraphics[width=0.45\linewidth]{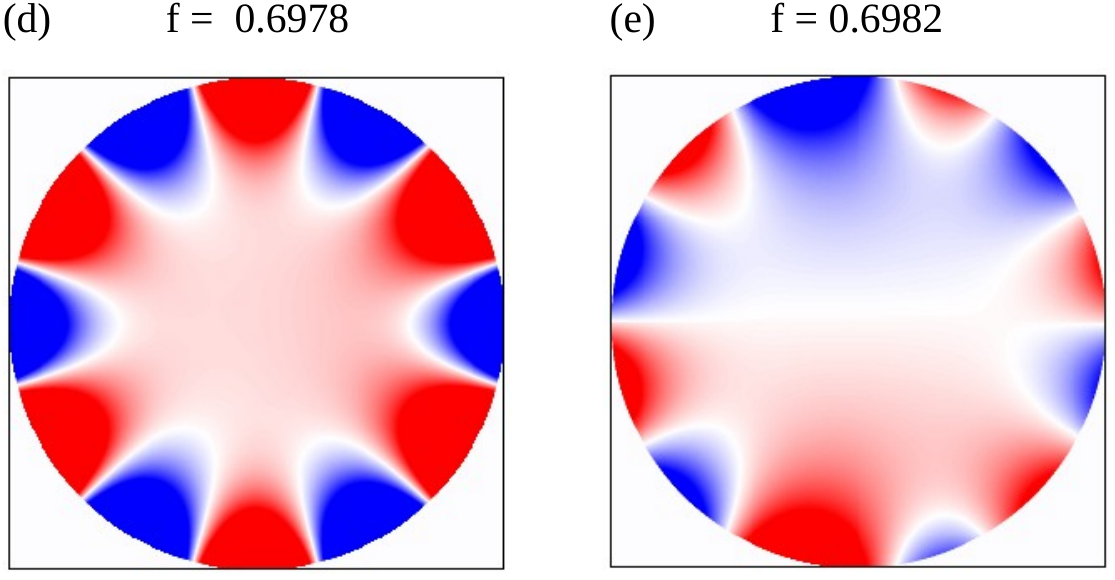}
\caption{(Color online)
The 6th bands. Left panel shows (a) the transmission of EM wave through linear array of metallic rods, (b) the transmittance
through 24 rows of rods, (c) the symmetric and asymmetric frequency bands. The magnetic field at frequency
$ f = 0.6978$ (lower symmetric band) and at the frequency $f = 0.6982$ (upper asymmetric band) is shown in right panels (d) and (e).
(The meaning of colors is given in caption to  Fig. 6.)
The lower (even) band $6+$ is well pronounced. The higher (odd) band is less visible since the incident angle was very small
($\theta = \pi/100$).}
\label{beta-6}
\end{figure}

To demonstrate behavior of the higher order resonances, we show in Figs.~\ref{beta-5} and \ref{beta-6} numerical results for the 5th and 6th bands.
We observed in the frequency range $0.692 < f  < 0.6945$  a doubly degenerate $E(10)$ state at the $\Gamma$-point which splits into a symmetric $A_1(10)$
and an asymmetric $B_2(10)$ state along the $\Gamma-X$ direction in the first  Brillouin zone -- see Fig.~\ref{beta-5}(c).
The lower band can be assigned to a symmetric 5- state belonging to the resonance of $\beta^{5-}$ coefficient shown in Fig.~\ref{koeficienty}(b)
which is associated with Fano resonance in the transmittance of the linear chain shown in Fig.~\ref{beta-5}(a).
The upper band can be assigned to an asymmetric 5+ state belonging to the resonance of $\beta^{5+}$ coefficient shown in Fig.~\ref{koeficienty}(b).
The field distribution associated with $A_1(10)$  band along the $\Gamma-X$ direction in the frequency range $0.692 < f  < 0.6935$  indicates symmetry corresponding to
$\sin 5\theta$  that resembles a symmetric 5- state -- see Fig.~\ref{beta-5}(d), while the field distribution associated with the higher band $B_2(10)$  in the frequency
range $0.6935 < f  < 0.6945$ reveals the symmetry $\cos 5\theta$ that resemble an asymmetric 5+  state -- see Fig.~\ref{beta-5}(e).
We note that due to the strongly localized distribution which is confined in the vicinity of cylinder surface, the frequency bands are very narrow.
Therefore, the effective refractive index of the structure \cite{sakoda} is very small. Consequently, transmission coefficient exhibits profound Fabry-Perot
oscillations and typically possesses rather small values even for symmetric bands -- see Fig.~\ref{beta-5}(b).

\begin{figure}[t!]
\includegraphics[width=0.80\linewidth]{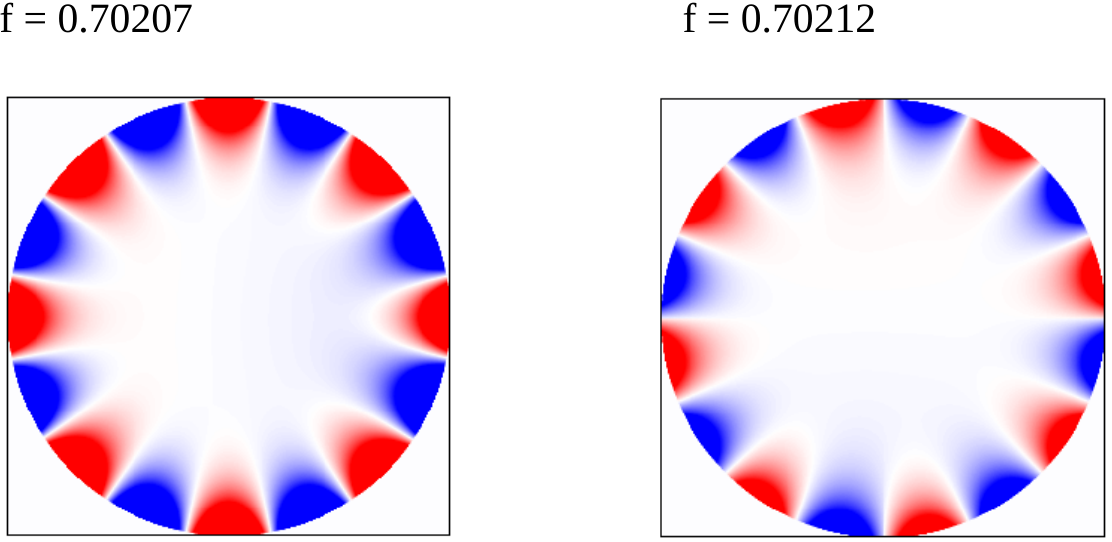}
\caption{Field distribution for even resonance  8+ (left, $|H_z|\le 10$ )  and odd resonance
 8- (right, $|H_z|\le 0.2$).  
The scale of the field reflects the  symmetry of bands
 and their coupling to the incident plane wave.
(The meaning of colors is given in caption to  Fig. 6.)
Note the change of the symmetry of the field (the incident plane wave propagates from the left to the right).
}
\label{beta-8}
\end{figure}

In the frequency range $0.6976 < f  < 0.6982$ we observed along the $\Gamma-X$ direction in the first Brillouin zone a asymmetric $B_1(12)$ and
an asymmetric $B_2(12)$ states \cite{EM}  -- see Fig.~\ref{beta-6}(c) -- which become degenerate at the $X$-point to form the $E(12)$ state. The lower band
can be assigned to a symmetric 6+ state belonging to the resonance of $\beta^{6+}$ coefficient
shown in Fig.~\ref{koeficienty}(b) which is associated with Fano resonance in the transmittance of the linear chain shown in Fig.~\ref{beta-6}(a).
The upper band can be assigned to an asymmetric 6- state belonging to the resonance of $\beta^{6-}$ coefficient shown in Fig.~\ref{koeficienty}(b).
The field distribution associated with $B_1(12)$  band along the $\Gamma-X$ direction in the frequency range $0.6976 < f  < 0.698$  indicates symmetry corresponding to
$\cos 6\theta$  that resembles a symmetric 6+ state -- see Fig.~\ref{beta-6}(d), while the field distribution associated with the upper band $B_2(12)$  in the frequency
range $0.698 < f  < 0.6982$ reveals the symmetry $\sin 6\theta$ that resembles an asymmetric 6-  state -- see Fig.~\ref{beta-6}(e).
The transmittance of the 2D structure shown in Fig.~\ref{beta-6}(a) displays strong oscillations in the frequency range corresponding to the symmetric band
and nearly invisible due to odd symmetry of the $B_2(12)$ band.

For the higher order resonances, the transmission bands are very narrow, as can be inferred from the shape of resonances shown in Fig.~\ref{koeficienty}(b) and (c).
Therefore, we do not display their transmission bands. Instead, we show in Fig.~\ref{beta-8} the field distribution associated for the 8th resonance. The change of the symmetry of the field with a very small variation of the frequency $\Delta f = 0.000070$ within the range $0.702050 < f < 0.702120$  demonstrate extremely narrow bandwidth of the higher order bands.

\section{Conclusions}
We numerically studied effects associated with a synergetic interplay between the microscopic and macroscopic resonance mechanisms
in the frequency and transmission spectrum of a two-dimensional array of metallic rods embedded in a vacuum. We have shown that the photonic
band structure consists of Bragg bands resulting from periodicity of the structure and Fano bands arising from
single scatterer Mie resonances. To explore the formation of Fano bands we studied a correspondence between series of flat bands and Fano resonances excited in a linear array of metallic rods by an incident $H$-polarized EM plane wave. We demonstrated that coupling between two underlying mechanisms affects the character of the band structure and depends strongly on the radius of the rod.  The symmetry of the modes obtained from numerical simulation has been determined by inspection of the distribution of the EM field associated with a selected cylinder in the 2D periodic structure considered. We have shown that the spatial distribution of the EM field displays characteristic patterns corresponding to the irreducible symmetry representations and reflects a different nature of both kinds of bands.

\section*{Acknowledgements}
This work was supported by the Slovak Research and Development Agency under the contract No. APVV-0108-11
and by the Agency  VEGA under the contract No. 1/0372/13. The research of V. Kuzmiak was supported by
Grant LD14028 of the Czech Ministry of Education within programme COST CZ.


\begin{thebibliography}{99}
\bibitem{fP} Our terminology does not holds for the lower part of the first frequency band, where the wavelength $\lambda\gg a$ and the 
wave propagation can be described within the effective permittivity model rather than in terms of theory of periodic media.
Since  the first band is never influenced by Fano resonances in our models, we do not discuss this frequency region throughout this paper.
\bibitem{note1} To assure the numerical stability, Hankel functions are  normalized by the frequency-dependent
function $H'_k(2\pi R/\lambda)$. 
\end{thebibliography}
\end{document}